\documentclass{article}

\usepackage{amsfonts}

\usepackage{graphicx}
\usepackage{amsmath}

\title{Motion of test bodies with internal degrees of freedom in non-Euclidian spaces.}
\author{J. J. S\L AWIANOWSKI \textdagger, B. GO\L UBOWSKA \textdaggerdbl \\
\it{Institute of Fundamental Technological Research,PAS, }\\
\it{21, \'{S}wi\c{e}tokrzyska str., 00-049 Warsaw, Poland} \\ e-mail: \textdagger jslawian@ippt.gov.pl,  \, \textdaggerdbl bgolub@ippt.gov.pl}

\begin{document}
\maketitle
\begin{abstract}
Discussed is mechanics of objects with 
internal degrees of freedom in generally 
non-Euclidean spaces. Geometric peculiarities 
of the model are investigated detailly. Discussed 
are also possible mechanical applications, e.g., 
in dynamics of structured continua,
defect theory and in other fields of mechanics 
of deformable bodies. Elaborated is a new method of 
analysis based on non-holonomic frames.
We compare our 
results and methods with those of other 
authors working in nonlinear dynamics (many 
of them refer to our papers \cite{Gol2004}, \cite{Gol2006}, \cite{IJJSco2004}, \cite{IIJJSco2005}). Simple examples 
of completely integerable models are presented.   
\end{abstract}
\section{Introduction}
In our earlier papers \cite{IJJSco2004}, \cite{IIJJSco2005} (and reference therein) we discussed systems with
collective and internal degrees of freedom ruled by the affine and linear groups,
first of all the metric-preserving groups, i.e., the isometry group and its homo-
geneous part, i.e., the rotation group (orthogonal group). Roughly speaking, we
were dealing there with rigid bodies, i.e., gyroscopes, and affinely-rigid bodies,
i.e., homogeneously deformable gyroscopes, both in the flat Euclidean space.
Such objects are interesting from the point of view of purely rational analytical
mechanics in itself, and besides, they occur in various quite practical problems
 \cite{Cas1995}, \cite{Rei1998}, \cite{Rei1996}, \cite{Ros1998}, \cite{Rub1985}, \cite{IJJSco2004}, \cite{IIJJSco2005}, \cite{Sousa1994}. In continuum mechanics they occur as
models of internal degrees of freedom, i.e., microstructure models in mechan-
ics of complex bodies. Classical examples are micropolar continua of brothers
Cosserat and micromorphic continua of Eringen \cite{Coss1909}, \cite{Coss1908}, \cite{2Coss1909}, 
\cite{Erin1968}, \cite{Erin1962}, \cite{Now1070}, \cite{Now1968}. Roughly speaking, they are continua of infinitesimal gyroscopes or
homogeneously deformable gyroscopes. They describe some granular media;
another application is the long-wave continuum limit of dynamics of molecular
crystals. Such continua do not consist any longer of the Newton type material
points; instead, their elementary constituents are material points with extra
attached orthonormal or general bases (respectively the micropolar and micromorphic continua). Let us mention there is also plenty of other microstructure
models where at material points some other geometric objects are attached like
e.g. liquid crystals which are, roughly speaking, continua of infinitesimal rods
(in continuous limit of course) \cite{Cap2003}, \cite{Cap1989}, \cite{Mar2000}, \cite{Yav2008}. To understand properly the dynamics of generalized continua, one must start from the well-defined dynamics
of extended affinely- rigid bodies or their systems in a flat Euclidean space.
\newline
Below we consider something more peculiar, namely the dynamics of very small,
essentially infinitesimal rigid or affinely-rigid bodies in a curved Riemannian
manifold, or even, more generally, in a manifold endowed only with affine connection. One must answer seriously the question concerning motivation. Such
models are mathematically very nice and interesting on the level of purely rational analytical mechanics. But it is clear that we are living in physical space
which in a very good approximation is Euclidean (in the scale of our every-day
life generally-realistic effects are negligible). Of course, one can answer immediately that this may be an intermediary step towards constructing relativistic
mechanics of continua \cite{Bress1978}, \cite{Kun1972},\cite{Sch1965},  \cite{Sop1975}. But topic, although interesting, is
perhaps esoteric from the point of view of every-day praxis. Let us notice, however that there exist some problems where the motion of small objects in curved spaces becomes a practically viable topic.
First of all let us observe that in a deformed elastic medium we are given two
spatial metric tensors: the usual metric tensor
$g_{ij}$
of the physical space, which
is in a sense the manifestation of the vacuum state of the gravitational field,
and the Cauchy deformation tensor $C_{ij}$, \cite{Erin1962}, \cite{Mars1983},
\begin{equation}
C_{ij}=\eta _{KL}\frac{\partial a^{K}}{\partial x^{i}}\frac{\partial a^{L}}{%
\partial x^{j}}.  \label{intr 1}
\end{equation}
In the above formula $\eta $ denotes the metric tensor of the material space
(the reference metric tensor), and $a^{K}$ are Lagrange (reference)
coordinates expressed as functions of the Euler (spatial) coordinates $x^{i}$.
Perhaps more naturally looks the expression for the reciprocal contravariant
tensor, 
\begin{equation}
C^{ij}=\frac{\partial x^{i}}{\partial a^{K}}\frac{\partial x^{j}}{\partial
a^{L}}\eta ^{KL},\qquad C^{im}C_{mj}=\delta ^{i}{}_{j}.  \label{intr 2}
\end{equation}
The point is that for small, concentrated objects in the deformed continuum,
like, e.g., defects (dislocations, disclinations, vacancies, interstitials \cite{Bloo1979}, \cite{Zor1967}), it
may be not $g_{ij}$ but just $C_{ij}$ that is felt as something like metric tensor. Similar
situations have place in solid state physics, where due to certain collective phenomena, it is used not the usual, geometric metric tensor, but rather the tensor
of effective mass which in certain situations even need not be positively definite.
It is reasonable to expect something similar here. Of course, the space endowed
with $C_{ij}$ as the metric tensor is still flat due to the non-separability (compatibility) conditions \cite{Erin1962}, \cite{Sop1975}, because the curvature tensor built of $C$ does vanish.
But it is well-known that the point defects feel the body manifold as curved
Riemannian space. Similar concepts appear in theory of residual stresses.
\newline
Second, let us put our attention on two-dimensional surface phenomena. The
surfaces of real bodies are curved submanifolds on dimension two. One can
expect that for many surface phenomena it is just the two-dimensional induced
metric that is relevant for the dynamics. For example, the microstructure of
the bulk of the body may generate some effective microstructure on the boundary, and no doubt that for the corresponding surface phenomena the induced
metric may be more relevant that the true metric tensor $g_{ij}$. Even without any
microstructure, the induced two-dimensional metric seems to be an important
factor, e.g., for the surface waves. Certainly it is so in dynamics of the boundary
membranes (films) of biological cells.
\newline
Let as also stress some ecological problems, like the motion of pollution regions
(spots) on the oceanic surface after damages of tankers. This is just the classical example of the motion of (relatively) small two-dimensional drop over the
curved spherical surface. One can also think about the motion of continental
plates and similar geophysical problems \cite{Gol2004}, \cite{JJSco2004}, \cite{IJJSco2004}.
\newline
Of course, it is convenient to begin any analysis from the motion of bodies in
constant curvature space, like the spherical or hyperbolic (Lobachevski) space,
because one can expect rigorous analytical solutions, just in virtue of the exis-
tence of high symmetries. There are some advantages of beginning the analysis
from the academic case of $n$-dimensional spaces restricting only then the considerations to physical values of $n$.
\newline
Within generally relativistic context the motion of extended structured bodies
was studied by Mathisson, Weyssenhoff, Tulczyjew and Papapetru \cite{51Math}, \cite{Tul1962},
\cite{Papa1951}; let us stress the contribution of the first three Polish physicists. On the
non-relativistic level the problems of rigid motion in curved spaces were discussed by Yugoslavian mechanician Stojanowi\u{c} \cite{Stoj1957}. We follow general ideals
of K\"{u}nzle \cite{Kun1972}, \cite{Sch1965} for whom such an object was described by linear frame attached to the material point. This is a mathematical idealization of very small
objects when the linear size of the body is small in comparison with the size
at which the spatial curvature changes remarkably. What we do is somehow
related to the pole-dipole approximation due to Mathisson, Weyssenhoff, Tulczyjew and Papapetrou. It is reasonable to expect that performing some more
detailled analysis one can obtain more adequate higher multipole description
where equations of motion contain not only Riemann tensor but also its covariant derivatives. Therefore, the situation will be particularly simple in symmetric
spaces (where the covariant derivative of the curvature tensor vanishes) \cite{Kob1963}. The
analysis probably would have to be lead with the help of normal coordinates
and make use of certain asymptotic power series expansion.
\newline
Before going any further, let us begin with explaining why we use here the approach based on the concept of infinitesimal rigid or affinely rigid bodies, at
least as a first approximation.
\newline
For extended systems of material points in a flat space, any two admissible
configurations of the usual rigid body (gyroscope) are related to each other by
some isometry transformation. The isometry group of the flat Euclidean space
of dimension $n$ is a $\frac{1}{2}n(n + 1)$- dimensional Lie group. However, in a generic
Riemann space the typical situation is that the isometry group has dimension
zero and consists of merely identity transformations. In any case this dimension
$k$ always satisfies the inequality
\begin{equation}
k\leq \frac{1}{2}n(n+1)  \label{intr 3}
\end{equation}
and the maximal possible value is attained in constant-curvature-spaces, including of course the flat spaces as a special example \cite{Kob1963}. But even in a non-flat
constant curvature space there are some problems, namely the centre of mass
concept is not correctly defined \cite{Shch2000}, \cite{Ste2000}, and because of this there is no splitting
into translational and internal motion. It is so when dealing with extended bodies. So there is only one way to escape the problem: to concentrate the attention
on sufficiently small bodies and go to the limit with their size. Then we obtain
just the non-Newtonian material point with something attached to it. In our
models this "something" is an orthonormal basis (infinitesimal gyroscope) or
just a general basis (infinitesimal affinely-rigid body, i.e., an infinitesimal homo-
geneously deformable body). Then mathematically everything becomes correct,
and physically one obtains a good approximate description of small metrically-
rigid or affinely-rigid bodies.
\newline
To be complete with all these arguments, let us remind the concepts of isome-
tries in Riemann spaces and affine transformations in spaces endowed with affine
connection. Let $(M,g)$ be an $n$-dimensional Riemannian space; $M$
is a differential manifold and $g$ the metric tensor on $M$. We say that the
mapping $\varphi $ of $M$ onto $M$ is an isometry if 
\begin{equation}
g=\varphi ^{\ast }g.  \label{intr 4}
\end{equation}
Analytically, when $\varphi $ is given by the dependence of the new point
coordinates $y^{i}$ on the old ones $x^{i}$, this is given by 
\begin{equation}
g_{ij}=g_{ab}\frac{\partial y^{a}}{\partial x^{i}}\frac{\partial y^{b}}{%
\partial x^{j}}.  \label{intr 5}
\end{equation}
\newline
Let now $\Gamma $ be an affine connection, e.g., the Levi-Civita one built
of $g$, but not necessarily, it may be a general connection, even completely
non-related to $g$. Let $\nabla $ \ denote the corresponding covariant
differentiation, and $\nabla _{X}$ the corresponding directional covariant
derivative along the vector field $X$. We say that the diffeomorphism $%
\varphi $\ of $M$ onto $M$ is an affine mapping if for any vector fields $X$%
, $Y$ on $M$ the following holds: 
\begin{equation}
\nabla _{\left( \varphi _{\ast }X\right) }\left( \varphi _{\ast }Y\right)
=\varphi _{\ast }\left( \nabla _{X}Y\right) ,  \label{intr 6}
\end{equation}
where $\varphi _{\ast }$ is an abbreviation for the action of $\varphi $ on
vector fields (thus, the covariant differentiation is transparent with
respect to the action of $\varphi $). Just like isometries, in a general
manifold with connection, the existence of affine diffeomorphisms is rather
exceptional.
\bigskip
\section{Degrees of freedom, kinematics, phase space, symmetries}
\bigskip
From now on the physical space $M$ is a differential manifold endowed with a
Riemannian structure, i.e., with some positively definite metric tensor
field $g$, or, in certain situations, even with the weaker structure given
by some affine connection $\Gamma $. One can also consider the double
structure when geometry of $M$ is given by some pair $\left( g,\Gamma
\right) $ where some relationship between $g$ and $\Gamma $ may exist or
not. As stated in Introduction, we replace extended bodies by structured
material points with attached linear bases. When those bases are by
definition $g$- orthonormal, we deal with the infinitesimal gyroscope; if
they are general, our system is an infinitesimal affinely rigid body
(homogeneous deformable gyroscope). One can also consider intermediate situations when some weaker constrains are imposed onto
infinitesimal affine motion (incompressibility etc.). The attached bases
describe internal degrees of freedom and give the symbolic description of
relative degrees of freedom after the limit transition, when the diameter of
the body tends to zero. Let our object be instantaneously placed at the
position $x\in M$ and some linear frame $e=(e_{1},\dots,e_{A},\dots,e_{n})$ be
attached at that point, thus 
\begin{equation}
e_{A}\in T_{x}M,\qquad A=1,\dots ,n.  \label{degrees 7}
\end{equation}
\newline
For infinitesimal affinely rigid bodies those are general linear frames, for
infinitesimal gyroscopes they are orthonormal with respect to the metric $g$, 
\begin{equation}
\left( e_{A}|e_{B}\right) =g\left( e_{A},e_{B}\right)
=g_{ij}e^{i}{}_{A}e^{j}{}_{B}=\eta _{AB},  \label{degrees 8}
\end{equation}
where $\eta $ is a fixed reference metric in the micromaterial space $N$.
Usually, but not always and not necessarily we put $N=R^{n}$ and take $\eta $
to be the Kronecker delta, 
\begin{equation}
\eta _{AB}=\delta _{AB}.  \label{degrees 9}
\end{equation}
\newline
The small latin indices refer to the physical space as seen, whereas the
capital ones refer to the micromaterial space $N$. Injections of the body
into M, or rather ino tangent spaces $T_{x}M$ are given by: 
\begin{equation}
y^{i}\left( t,a\right) =x^{i}\left( t\right) +\varphi ^{i}{}_{K}\left(
t\right) a^{K},  \label{degrees 10}
\end{equation}
where $x^{i}$ is the position of the material point, the quantities $\varphi ^{i}{}_{K}$
describe the relative/internal motion, and $a^{K}$ \ are Lagrangian
coordinates in the micromaterial space $N$. 
More precisely, the formula (\ref
{degrees 10}) is an approximation which is valid in virtue of the
infinitesimal character of the body; the better valid, the smaller is the body. One can imagine it as formulated in terms of some normal coordinates in an $\varepsilon $-order size neighborhood of the point $x\in M$, when $\varepsilon $ tends to zero.
\newline
The configuration space of an infinitesimal affine body is given by the
principal fibre bundle manifold of linear frames $FM$, or to be more
precise, by the connected component of $FM$, to exclude the nonphysical
singular configurations, when the system $e=(e,\dots,e_{A},\dots,e)$\ fails to
be linearly independent. According to the ideas of differential geometry,
interpretation of $FM$ as a principal fibre bundle means that the group $%
{\rm GL}\left( n,R\right) $ acts freely and transitively on this manifold \cite{Kob1963}, 
\begin{equation}
e=(e,\dots,e_{A},\dots e)\rightarrow eL= (\dots,e_{B}L^{B}{}_{A},\dots )
\label{degrees 11}
\end{equation}
(do not confuse this action with the left hand side action of ${\rm GL}\left(
T_{x}M\right) $ on $T_{x}M$ and the manifold of bases in $T_{x}M$). More
precisely, the subgroup of positive-determinant matrices ${\rm GL}^{+}\left(
n,\mathbb{R}\right) \subset {\rm GL}\left( n,\mathbb{ R}\right) $ acts freely and transitively on the
connected component of $FM$.
\newline
Equivalently, instead $FM$ one can use the manifold of co-frames $F^{\ast }M$,
i.e., frames of covariant vectors (linear functions on $T_{x}M$ at all
possible $x\in M$). There is an obvious natural diffeomorphism between $FM$
and $F^{\ast }M$ given by the duality mapping 
\begin{equation}
FM\ni e=\left( \dots ,e_{A},\dots \right) \rightarrow \widetilde{e}=\left(
\dots ,e^{A},\dots \right) , \label{degrees 12}
\end{equation}
\ where 
\begin{equation}
e^{A}\left( e_{B}\right) =\left\langle e^{A},e_{B}\right\rangle
=e^{A}{}_{i}e^{i}{}_{B}=\delta ^{A}{}_{B}.  \label{degrees 13}
\end{equation}
Here the group ${\rm GL}\left( n,R\right) $ acts on the left according to the
rule: 
\begin{equation}
\widetilde{e}=\left( \dots ,e^{A},\dots \right) \rightarrow
\widetilde{e}L=
(\dots ,L^{-1A}{}_{B}e^{B},\dots ).  \label{degrees 14}
\end{equation}
If $\dim M=n$ then $\dim FM=\dim F^{\ast }M=n(n+1)$ ; this is the number of
degrees of freedom.
\newline
Any system of local coordinates $x^{i}$ on $M$ gives rise to the obvious
coordinates $(x^{i},e^{i}{}_{A})$ on $FM$ or $(x^{i},e^{A}{}_{i})$ on.$%
F^{\ast }M$; here $e^{i}{}_{A}$ are components of $e_{A}$ with respect to $%
x^{i}$. For simplicity we do not distinguish graphically between $x^{i}$ and
their pull-backs $x^{i}\circ \pi $, $x^{i}\circ \pi ^{\ast }$ respectively to 
$FM$ and $F^{\ast }M$. Here $\pi :FM\rightarrow M$, $\pi ^{\ast }:F^{\ast
}M\rightarrow M$ are natural projections which assign to the frames $%
e_{x}\in F_{x}M\subset FM$, $\widetilde{e}_{x}\in F^{*}_{x}M\subset F^{*}M$
their attachment points $x\in M$
\newline
If $M$ is an affine (flat) space with the linear space of translations $V$,
then obviously $FM$, $F^{*}M$ are canonically diffeomorphic respectively
with $M\times F(V)$, $M\times F(V^{*})$; $F(V)$ and $F(V^{*})$ denote here
the manifolds of frames in $V$ and $V^{*}$ (more rigorously - their connected
components). This is the byproduct of the fact that the tangent and
cotangent bundles $TM$, $T^{*}M$ are isomorphic respectively with the
Cartesian products $M\times V$, $M\times V^{*}$. Then the degrees of freedom
split naturally into translational ($M$) and internal ($V$) ones. If $M$
is a curved manifold, this is no longer the case; there is no canonical
identification with Cartesian products, and often there is no
identification at all because of topological obstacles. But of course,
translational motion in $M$ is always well defined due to the projection $%
\pi : FM \rightarrow M$. If the total motion in $FM$ (always well defined)
is given by a curve $\varrho :\mathbb{R}\rightarrow FM$ ($\mathbb{R}$ is the
time axis), then the translational motion is described by the projected
curve $\pi \circ \varrho :\mathbb{R}\rightarrow M$. Translational velocity
is also well defined and given at the time instant $t\in \mathbb{R}$ by the
tangent vector 
\begin{equation}  \label{degrees 15}
v(t)=(\pi \circ \varrho )^{^{\prime}}(t)= \left( T \pi \circ \varrho
^{^{\prime}} \right) (t)\in T _{\pi (\varrho (t)) } M.
\end{equation}
Unlike this, neither the internal motion nor internal velocity are well
defined in a bare structure- less manifold. If our internal degrees of
freedom are constrained to gyroscopic ones, the configuration space becomes $%
(FM,g)$, the manifold of $g$- orthonormal frames. Then the number of degrees
of freedom becomes reduced to 
\begin{equation}\label{degrees 16}
\dim (FM,g)=\frac{1}{2} n(n+1),
\end{equation}
i.e., $n$ translational degrees of freedom, parametrized by spatial
coordinates $x^{i}$ and 
\begin{equation}  \label{degrees 17}
\frac{1}{2}n(n-1)
\end{equation}
internal gyroscopic deegres of freedom, in a sense like in the flat
Euclidean space. In the latter case the configuration space would be
identifiable with 
\begin{equation}\label{degrees 18}
M\times F(V,g),
\end{equation}
or rather with its connected components; $F(V,g)$ denotes here the
submanifold of $g$- orthonormal linear frames in $V$. Some important
problem appears here, which is still important when the full system of
affine degrees of freedom is concerned. Namely, in virtue of constrains (\ref
{degrees 8}) the quantities $x^{i}, e^{i} {}_{A}$ fail to be
functionally independent and they are not generalized coordinates any
longer. In a flat space one can help with the problem of defining independent
coordinates in the following way: one fixes some particular orthonormal
reference frame $E$ in $M$ and then "parametrizes" $F(V,g)$ via 
(\ref{degrees 11}), using orthogonal matrices $L\in SO(V,g)$. Those in turn are parametrized
in any of well- known ways, e.g., using canonical coordinates of the first
kind, Euler angles etc. How to follow this pattern in a curved manifold? The
only natural way is to introduce some non-holonomic orthonormal reference
reference frame $E$ in $M$, i.e., some auxiliary field 
\begin{equation}  \label{degrees 19}
E=(\dots ,E_{A},\dots )
\end{equation}
on $M$, where 
\begin{equation}  \label{degrees 20}
(E_{A}|E_{B})=g(E_{A},E_{B})=g_{ij}E^{i}{}_{A}E^{j}{}_{B}=\eta_{AB}
\end{equation}
and usually we put $\eta _{AB}=\delta _{AB}$, obviously when $g$ (as
assumed) is positively definite. If $(M,g)$ is curved (Riemannian), then $E$
must be non-holonomic.
\newline
Let us remind that in differential geometry the non-holonomy object $\Omega $
of a field of frames $E$ is defined by any of the following equivalent
formulas \cite{Zor1967}: 
\begin{equation}\label{degrees 21}
[ E_{A},E_{B}]=\Omega ^{C}{}_{AB}E_{C},\qquad dE^{A}=\frac{1}{2}\Omega
^{A}{}_{BC}E^{C} \wedge E^{B}, 
\end{equation}
where, as usual $E^{A}$ are elements of the co-frame $\widetilde{E}$ dual to 
$E$, ''$d$'' denotes the exterior differentiation, and for any vector fields 
$X$, $Y$ their Lie bracket $[X,Y]$ is analytically expressed by
\begin{equation}\label{degrees 22}
[X,Y]^{i}=X^{j}\frac{\partial Y^{i}}{\partial x^{j}}-Y^{j}\frac{\partial
X^{i}}{\partial x^{j}};
\end{equation}
\newline
this vector field is correctly defined, i.e., it does not 
depend on the choice of coordinates $x^{i}$ \cite{Kob1963}. The object $\Omega $ 
vanishes if and only if $E$ is holonomic, i.e., consist of 
vector fields tangent to some coordinate lines. The corresponding 
tensor field on $M$,
\begin{equation}\label{degrees 23}
S[E]=\frac{1}{2}\Omega ^{A}{}_{BC}E_{A}\otimes \widetilde{E}^{B}\otimes 
\widetilde{E}^{C},
\end{equation}
i.e., analytically
\begin{equation}\label{degrees 24}
S^{i}{}_{jk}=\frac{1}{2} E ^{i}{}_{A}\left(\frac{ \partial }{\partial x^{k}} E^{A}{}_{j} - \frac{ \partial }{\partial x^{j}} E^{A}{}_{k}\right),
\end{equation}
is the torsion of the teleparallelism connection $\Gamma _{{\rm tel}}[E]$ built of $E$,
\begin{equation}\label{degrees 25}
S^{i}{}_{jk}=\frac{1}{2} \left({\Gamma_{{\rm tel}}[E]}^{i}{}_{jk} - {\Gamma_{{\rm tel}}[E]}^{i}{}_{kj}\right),
\end{equation}
The connection $\Gamma [E]$ is analytically given by
\begin{equation*}
{\Gamma _{{\rm tel}}[E]}^{i}{}_{jk}=E^{i}{}_{A}\frac{\partial}{\partial x^{k}}E^{A}{}_{j}.
\end{equation*}
Geometrically it is uniquely defined by the
demand that the vector fields $E_{A}$ are all parallel 
with respect to the corresponding affine connection,
\begin{equation}\label{degrees 26}
\overset{E}{\nabla}  E_{A}=0, \qquad A=1,\dots ,n.  
\end{equation}
When $E$ is non-holonomic, the torsion of $\Gamma [E]$ does 
not vanish, but the curvature tensor $\mathcal{R}[E] $ of any $\Gamma [E]$ 
is always vanishing. Incidentally, let us mention
 that the teleparallelism connection is directly
 related to the non-holonomic representation of any
 possible affine connection. Namely, if $\Gamma$ is some
 affine connection on $M$, then its non -holonomic
representation with respect to $E$ is given by
coefficients $\Gamma ^{A}{}_{BC} $ such that
\begin{equation}\label{degrees 27}
\nabla _{C} E_{B}= \Gamma ^{A}{}_{BC}E_{A};
\end{equation}
here $\nabla $ denotes the covariant differentiation in the 
$\Gamma $-sense. One can easily show that
\begin{equation}\label{degrees 28}
\Gamma ^{i}{}_{jk} - \Gamma _{{\rm tel}}[E]^{i}{}_{jk}=E^{i}{}_{A}\Gamma ^{A}{}_{BC}E^{B}{}_{j}E^{C}{}_{k}.
\end{equation}
In particular,
\begin{equation}\label{degrees 29}
\Gamma _{{\rm tel}}[E]^{A}{}_{BC}=0;
\end{equation}
obviously, we mean here non-holonomic coefficients
of $\Gamma _{{\rm tel}}[E] $ with respect to $E$ itself. 
\newline
Let us go back to the main problem of our 
non-holonomic description. Assume that at some time 
instant $t\in \mathbb{R} $ the structured material point is placed
at $x(t)\in M $ and its internal degrees of freedom 
have the attitude
\begin{equation}\label{degrees 30}
e(t)=(\dots ,e_{A}(t),\dots ) \in F_{x(t)}M; \qquad e_{A}(t)\in T_{x(t)}M.
\end{equation}
The reference frame just passed then is given by 
\begin{equation}\label{degrees 31}
E_{x(t)}=(\dots ,E_{A}{}_{x(t)},\dots )\in F_{x(t)}M; ; \qquad E_{A}{}_{x(t)}\in T_{x(t)}M.
\end{equation}
Being attached at the same point $x(t)\in M$ the 
vectors $e_{A}(t)$ may be expanded with respect to the
frame $E_{x(t)}$,
\begin{equation}\label{degrees 32}
e_{A}(t)=E_{B}{}_{x(t)}L^{B}{}_{A}(t).
\end{equation}
In this way, if $E$ is globally defined and kept fixed,
one can interpret the curve
\begin{equation}\label{degrees 33}
\mathbb{R} \ni t \rightarrow L(t) \in {\rm GL} (n,\mathbb{R} )
\end{equation}
as a correct description of the internal part of motion, and 
the configuration space $FM$ (its connected component) 
may be identified with the Cartesian product 
\begin{equation}\label{degrees 34}
M \times {\rm GL}(n,\mathbb{R} )\qquad \left( M\times {\rm GL}^{+}(n,\mathbb{R} ) \right).
\end{equation}
This representation becomes particularly important 
when gyroscopic constraints are imposed. Then obviously 
the matrices $L(t)$ become orthogonal ($\eta$-orthogonal) 
and the configuration space is identified with
\begin{equation}\label{degrees 35}
M\times SO(n,\mathbb{R} ) \qquad \left( M\times SO(\eta )\right).
\end{equation}
Such description is very important technically, 
because, just like in the flat space model, the group
$SO(n,\mathbb{R}) $ (in particular $SO(2,\mathbb{R}) $, $SO(3,\mathbb{R}) $ may be
parametrized in a variety of standard ways, and in 
this way some good, non-redundant, generalized coordinates
may be constructed.
\newline
But even much more: Even without gyroscopic 
constraints, generalized coordinates $(x^{i}, e^{i}{}_{A}) $ which are
well-defined then, are non-useful in realistic problems.
Unlike this, the representation (\ref{degrees 34}) is well-suited
to various viable dynamical models. The idea is to use
the polar decomposition of $L$ or its two -polar decomposition 
(singular value decomposition) \cite{JJSco2004}, \cite{IJJSco2004}, \cite{IIJJSco2005}. In resulting coordinates 
some physically viable dynamical models are analytically 
treatable. Because of this the description based on the 
use of auxiliary non-holonomic frames $E$ is both technically 
useful and geometrically interesting. It enables one 
to reduce many curved -space- formulaes to ones 
formally very similar to expressions appearing in 
mechanics of extended bodies in flat spaces \cite{Gol2004}, \cite{Gol2006}. 
As mentioned, in a bare structureless manifold $M$ there are two 
kinds of well-defined velocities: generalized velocity of the 
total motion in $FM$ (or $(FM,g)$), $\rho '(t) $ parametrized by 
$\left( \dots ,\frac{dx^{i}}{dt},\dots ,\frac{d}{dt}e^{i}{}_{A},\dots  \right) $ and translational motion velocity
$v(t)=(\pi \circ \rho)'(t) $, i.e., (\ref{degrees 15})  parametrized by $\left( \dots ,\frac{dx^{i}}{dt},\dots \right)$. 
The quantities $\frac{d}{dt}e^{i}{}_{A}$ fail to be well-defined internal 
velocities;moreover, they are not components of vectors in $M$. 
If some reference frame $E$ is fixed, the time
derivatives of $L$-matrices with components
\begin{equation}\label{degrees 36}
\mathcal{V}({\rm rl})^{A}{}_{B}(t):=\frac{d}{dt} L^{A}{}_{B}(t)
\end{equation}
are well--defined measures of internal velocity. 
Usually it is more convenient to use Lie-algebraic objects 
denoted by $\Omega _{{\rm rl}}$, $\widehat{\Omega}_{{\rm rl}} $ and given by formulas:
\begin{equation}\label{degrees 37}
\Omega_{{\rm rl}}{}^{A}{}_{B}=\left( \frac{d}{dt}L^{A}{}_{C}\right)L^{-1}{}^{C}{}_{B},
\end{equation}
\begin{equation}\label{degrees 38}
\widehat{ \Omega}_{{\rm rl}}{}^{A}{}_{B}=L^{-1}{}^{A}{}_{C} \frac{d}{dt}L^{C}{}_{B}= L^{-1}{}^{A}{}_{C}\Omega _{{\rm rl}}{}^{C}{}_{D}L^{D}{}_{B}.
\end{equation}
Obviously, they are non--holonomic velocities, i.e., they are not 
time derivatives of any generalized coordinates. The 
label (rl) means relative; they describe internal motion 
with respect to the fixed reference frame $E$. Because of this 
essential dependence on the calculation-helping auxiliary 
quantity they are perhaps not very convincing, both physically 
and geometrically. Indeed $E$ does not describe any real geometry. And it 
would be rather difficult to construct convincing models 
of kinetic energy (Riemannian geometry of the configuration 
space) using just these quantities and basing on the 
analogy with motion of extended bodies in flat space. 
Much more natural measure of internal motion velocities 
is based on covariant derivatives, because the affine 
connection $\Gamma ^{i}{}_{jk} $ is a "real" geometry of $M$, not an
auxiliary analytical tool. Usually, although not necessarily,
$\Gamma ^{i}{}_{jk}$ is the Levi-Civita connection $\{^{i} _{jk}  \} $ built of the
metric tensor $g_{ij}$ on $M$:
\begin{equation}\label{degrees 39}
\Gamma ^{i}{}_{jk}=\{ ^{i} _{jk} \}= \frac{1}{2}g^{im} \left( g_{mj,k}+g_{mk,j} - g_{jk,m} \right).
\end{equation}
Let us remind that $\{ ^{i} _{jk} \} $ is uniquely defined by the 
condition that the metric tensor is parallel,i.e., that its 
covariant derivative with respect to $\{^{i} _{jk} \} $ vanishes,
\begin{equation}\label{degrees 40}
\overset{\Gamma}{ \nabla_{k}} g_{ij}=0,
\end{equation}
and, in addition, it is symmetric, i.e., torsion-free:
\begin{equation}\label{degrees 41}
\Gamma^{i}{}_{jk}=\Gamma^{i}{}_{kj}.
\end{equation}
In certain problems it may be useful to admit a 
weaker relationship between affine connection and metric 
structures, e.g., Riemann-Cartan space, Weyl space, 
Riemann--Cartan--Weyl space, or no relationship at all. 
There are also interesting models when only affine 
connection but no metric structure is assumed in $M$.
\newline
Affine connection enables one to define vectors of 
internal velocities $V^{i}{}_{A}$:
\begin{equation}\label{degrees 42}
V^{i}{}_{A}=\frac{D}{Dt}e^{i}{}_{A}=\frac{d}{dt}e^{i}{}_{A}+\Gamma ^{i}{}_{jk}\left(x(t)\right) e^{j}{}_{A}(t)\frac{dx^{k}}{dt}.
\end{equation}
This is simply the covariant differentiation of 
the attached vectors $e_{a}$ along the curve of translational 
motion in $M$; the latter is described by the time 
dependence of spatial coordinates $x^{i}(t)$. When $(M,\Gamma )$ 
is a flat affine manifold (vanishing curvature and 
torsion tensors), then obviously, $FM$ trivializes to 
$M\times F(V)$ and the quantities $\frac{D}{Dt}e^{i}{}_{A}$ reduce in 
affine coordinates to usual derivatives $\frac{d}{dt}e^{i}{}_{A}$. In a 
general manifold with non-flat affine connection, 
the above quantities $V^{i}{}_{A}$ (\ref{degrees 42}) are non-holonomic 
velocities, i.e., they are not time derivatives of 
generalized coordinates. And the total covariant velocity 
$\left(\dots ,v^{i},\dots ;\dots ,V^{i}{}_{A},\dots \right)$ with components
\begin{equation}\label{degrees 43}
\left(\dots ,v^{i},\dots ;\dots ,V^{i}{}_{A},\dots \right)=\left(\dots ,\frac{dx^{i}}{dt},\dots ; \dots ,\frac{D e^{i}{}_{A}}{Dt},\dots \right)
\end{equation}
is non holonomic in this sense. 
Let us now express these velocities in terms of 
the description based on the use of auxiliary 
(usually g-orthonormal) reference field of frames $E$.
\newline
For any virtual motion, not necessarily one satisfying any
equations of motion we have in virtue of (\ref{degrees 32})
\begin{equation}\label{degrees 44}
V_{A}=\frac{D}{Dt}e_{A}=\left(\frac{D}{Dt}E_{x(t)}{}_{B}\right) L^{B}{}_{A}(t) + E_{x(t)}{}_{B}\frac{D}{Dt} L^{B}{}_{A}(t).
\end{equation}
Obviously, from the point view of geometry of $M$, 
$L^{B}{}_{A}(t)$ are scalar quantities, thus their covariant 
derivatives reduce to the usual ones,
\begin{equation}\label{degrees 45}
\frac{D}{Dt}L^{B}{}_{A}(t)=\frac{d}{dt}L^{B}{}_{A}(t).
\end{equation}
The projected curve $\mathbb{R}\ni t \rightarrow x(t)$ describes translational 
motion in $M$ (analytically, $x^{i}(t)$), and the first
factor in the first term of (\ref{degrees 44}) will be expressed 
by the $\Gamma$-covariant derivative of the field $E$:
\begin{equation}\label{degrees 46}
\frac{D}{Dt}E_{x(t)}{}_{B}=\left(\nabla _{i}  E_{B}\right) _{x(t)} \frac{dx^{i}}{dt}= \left(\nabla _{C}  E_{B}\right) _{x(t)} E^{C}{}_{i}{}_{x(t)}\frac{dx^{i}}{dt}
\end{equation}
In analogy to the flat-space theory of extended bodies 
we shall use affine velocities, i.e., what Eringen used 
to call gyration \cite{Erin1968} in this theory of micromorphic
continua,
\begin{equation}\label{degrees 47}
\Omega^{i}{}_{j}=\left( \frac{D}{Dt} e^{i}{}_{A}\right) e^{A}{}_{j} , \qquad \widehat{\Omega}^{A}{}_{B}=e^{A}{}_{i}\frac{D}{Dt} e^{i}{}_{B}.
\end{equation}
Roughly speaking, this is respectively spatial and 
co-moving representation of the same quantity,
\begin{equation}\label{degrees 48}
\Omega^{i}{}_{j}=e^{i}{}_{A}\widehat{\Omega}^{A}{}_{B}e^{B}{}_{j}.
\end{equation}
If motion is $g$-rigid, i.e., the frames $e_{A}$ are 
constrained to be orthonormal, 
these tensors are skew--symmetric with respect to the corresponding 
spatial and material metrics,
\begin{eqnarray}\label{degrees 491}
&&\Omega ^{i}{}_{j}= -\Omega _{j}{}^{i}=-g_{jk}g^{il}\Omega^{k}{}_{l}   \\
&& \widehat{\Omega}^{A}{}_{B}= -\widehat{\Omega}_{B}{}^{A}=-\eta_{BC}\eta^{AD}\widehat{\Omega}^{C}{}_{D} \label{degrees 492}
\end{eqnarray}
and become the usual angular velocities. In exceptional but
physical dimension $n=3$ they are identified (as all skew--symmetric 
tensors) with axial vectors referred to as angular velocity 
vectors.
\newline
We shall use the following notation for objects appearing in
formulas (\ref{degrees 44}-\ref{degrees 48}):
\begin{eqnarray}\label{degrees 50}
 \widehat{\Omega }_{{\rm rl}}{}^{A}{}_{B}&=& L^{-1}{}^{A}{}_{C} \frac{d}{dt} L^{C}{}_{B}, \\\label{degrees 51}
\widehat{\Omega }_{{\rm dr}}{}^{A}{}_{B} &=& L^{-1}{}^{A}{}_{F}\Gamma^{F}{}_{DC}L^{D}{}_{B}L^{C}{}_{G}v^{G},
\end{eqnarray}
where $v^{G}$ are co-moving components of translational 
velocity vector,
\begin{equation}\label{degrees 52}
v^{G}=e^{G}{}_{i}\frac{dx^{i}}{dt}
\end{equation}
The meaning of labels "rl", "dr" is respectively "relative" 
and "drive".$ \widehat{\Omega }_{{\rm rl}}$ describes the affine velocity with respect 
to the just passed reference frame $E_{x(t)}\in F_{x(t)}M$. 
$\widehat{\Omega} _{{\rm dr}}$ is the affine velocity (angular velocity and strain rate)
of the frame $E$ itself as seen by the moving observer. This 
quantity describes the contribution of translational velocity
to the total $\widehat{\Omega}^{A}{}_{B}$. This method of non-holonomic frames 
was intensively used by M. Z\'{o}rawski in his book and papers  
on defect theory; \cite{Zor1967} and references therein. 
\newline
Canonical momenta conjugate to generalized coordinates
$x^{i}$, $\varphi ^{i}{}_{A}$ will be denoted by $p_{i}$, $p^{A}{}_{i}$. In many problems 
it is convenient to use non holonomic momenta conjugate 
to non-holonomic velocities $\Omega ^{i}{}_{j}$, $\widehat{\Omega}^{A}{}_{B}$, $\widehat{v}^{A}$. By analogy 
to mechanics of extended bodies in flat spaces we 
shall use the terms affine spin (hypermomentum) 
respectively in laboratory and co-mowing representation, 
denoted respectively by $\Sigma^{i}{}_{j}$, $\widehat{\Sigma}^{A}{}_{B}$. As mentioned, the 
covariant velocities of the total motion are given by
\begin{equation}\label{degrees 53}
\left(\dots ,v^{i},\dots ;\dots ,V^{i}{}_{A},\dots \right)= \left(\dots ,\frac{dx^{i}}{dt},\dots ;\dots ,\frac{D}{Dt}e^{i}{}_{A},\dots \right);
\end{equation}
if $\Gamma$ is non-Euclidean, they are non-holonomic,
\begin{eqnarray}\label{degrees 54}
V^{i}{}_{A}&=&v^{i}{}_{A}+\Gamma^{i}{}_{jk}e^{j}{}_{A}v^{k},\\
v^{i}{}_{A}&=&\frac{d}{dt}e^{i}{}_{A}.\label{degrees 55}
\end{eqnarray}
Their conjugate non-holonomic momenta $(p_{i},P^{A}{}_{i})$
satisfy:
\begin{equation}\label{degrees 56}
P_{i}=p_{i}-e^{j}{}_{A}P^{A}{}_{k}\Gamma^{k}{}_{ji},\qquad P^{A}{}_{i}=p^{A}{}_{i},
\end{equation}
and the spin quantities are given by
\begin{equation}\label{degrees 57}
\Sigma^{i}{}_{j}=e^{i}{}_{A}P^{A}{}_{j}, \qquad\widehat{\Sigma}^{A}{}_{B} = P^{A}{}_{i}e^{i}{}_{B}=e^{i}{}_{A}\widehat{\Sigma}^{A}{}_{B}e^{B}{}_{j}
\end{equation}
Let us observe the characteristic duality: $v^{i}$ is 
a well-defined vector in $M$, whereas $v^{i}{}_{A}=\frac{d}{dt}e^{i}{}_{A}$
fail to be so. Unlike this $P^{A}{}_{i}=p^{A}{}_{i}$ are components of
 well defined co-vectors in $M$, whereas $p_{i}$ themselves
do not represent co-vector components in $M$. The all 
dualities mentioned here and underlying the definitions 
of $p_{i}$, $P^{A}{}_{i}$, $\Sigma^{i}{}_{j}$, $\widehat{\Sigma}^{A}{}_{B}$ are meant in the sense:
\begin{eqnarray}\label{degrees 58}
p_{i}v^{i}+P^{A}{}_{i}V^{i}{}_{A}=P_{i}V^{i}+P^{A}{}_{i}V^{i}{}_{A}=
p_{i}v^{i}+\Sigma ^{i}{}_{j}\Omega ^{j}{}_{i}= \widehat{P} _{A}\widehat{V}^{A}+\widehat{\Sigma}^{A}{}_{B}\widehat{\Omega}^{B}{}_{A}
\end{eqnarray}
where
\begin{equation}\label{degrees 59}
\widehat{P}_{A}=P_{i}e^{i}{}_{A}.
\end{equation}
After some calculations one can show that the usual 
Poisson brackets for the phase space coordinates have the form 
\begin{eqnarray}\label{degrees 601}
\{x^{i},x^{j}\}= \{e^{i}{}_{A},e^{j}{}_{B} \}=0,\quad \{x^{i},e^{j}{}_{A}\}= \{P^{A}{}_{i},P^{B}{}_{j}\}=0,
\\\label{degrees 602}
\{P^{A}{}_{i},x^{j}\}=0,\quad \{e^{i}{}_{A},P^{B}{}_{j}\}=\delta ^{i}{}_{j} \delta ^{B}{}_{A},\quad \{x^{i},P_{j}\}=\delta ^{i}{}_{j}.
\end{eqnarray}
There is nothing surprising in those brackets;
they look exactly as ones with $P_{i}$ replaced by $p_{i}$;
(remember that $P^{A}{}_{i}=p^{A}{}_{i}$; cf. (\ref{degrees 56}). However,
(\ref{degrees 601}) (\ref{degrees 602}) is not a complete system and the Poisson brackets
like $\{P_{i},P_{j}\}$, $\{P_{i},P^{A}{}_{j}\}$, $\{P_{i},e^{j}{}_{A}\}$ are
different than their holonomic analogues, like, e.g.,
\begin{equation}\label{degrees 61}
\{p_{i},p_{j}\}=0,\qquad \{p_{i},p^{A}{}_{j}\}=0,\qquad\{p_{i},e^{j}{}_{A}\}=0,
\end{equation}
etc. Indeed, using the expressions (\ref{degrees 57}), one
can easily show that
\begin{eqnarray}
\{P_{i},P_{j}\}&=&\Sigma^{k}{}_{l}R^{l}{}_{kij},\nonumber
\\
 \{P_{i},P^{A}{}_{j} \}&=& -P^{A}{}_{k}\Gamma^{k}{}_{ji},\label{degrees 62}
\\ 
\{P_{i},e^{j}{}_{A}\}&=&e^{k}{}_{A}\Gamma ^{j}{}_{ki},\nonumber
\end{eqnarray}
where $R^{l}{}_{kij} $ is the curvature tensor in the convention
\begin{equation}\label{degrees 63}
R^{l}{}_{kij}=\Gamma ^{l}{}_{kj,i}-\Gamma ^{l}{}_{ki,j}+
\Gamma ^{l}{}_{ai} \Gamma ^{a}{}_{kj}-\Gamma ^{l}{}_{aj} \Gamma ^{a}{}_{ki}
\end{equation}
and comma denotes the partial derivative with respect
to the indicated coordinates $ x^{i}$.
Obviously, in a flat affine space, (\ref{degrees 62}) reduce to 
(\ref{degrees 61}).
The Poisson brackets (\ref{degrees 62}) mean geometrically that,
$P_{i}$ are Hamiltonian generators of the parallel 
transport of state variables. In a flat space they are
just Hamiltonian generators of spatial translations. It 
is easy to find another Poisson brackets, also convenient 
in applications, and at the same time admitting a clear 
geometric interpretation. For example,
\begin{equation} \label{degrees 64}
\{\Sigma ^{a}{}_{b},\Sigma ^{i}{}_{j}\}=\delta ^{a}{}_{j}\Sigma^{i}{}_{b} - \delta ^{i}{}_{b}\Sigma^{a}{}_{j}.
\end{equation}
Here we easily discover just the well-known structure 
constants of the linear group. And no wonder, $\Sigma^{k}{}_{l} $, 
which are referred to as affine spin (hypermomentum)
 components, are Hamiltonian generators of linear 
transformations of ${\rm GL}(T_{x}M)$ in $FT_{x}M$ (in the 
manifold of linear frames, i.e., internal degrees of freedom \cite{Abr1978}, \cite{Arn1978}, \cite{Bur1996}, \cite{Mars1999}, \cite{JJSco2004}, \cite{IJJSco2004}.
Let us remind, if we have the field of non-singular 
mixed tensors in $M$,
\begin{equation} \label{degrees 65}
M\in x \rightarrow L_{x} \in {\rm GL}(T_{x}M)\subset L(T_{x}M)\simeq T^{1} _{1}(T_{x}M),
\end{equation}
it acts on our configuration space, more exactly on 
the manifold of internal degrees of freedom, according 
to the rule
\begin{equation} \label{degrees 66}
F_{x}M\ni e=(\dots ,e_{A},\dots )\rightarrow  (\dots ,L_{x}\circ e_{A},\dots ),
\end{equation}
i.e., analytically:
\begin{equation}\label{degrees 67}
(\dots  ,x^{a},\dots ;\dots  ,e^{i}{}_{A},\dots  )\rightarrow  (\dots  ,x^{a},\dots  ;\dots  ,L^{i}{}_{j}(x) e^{j}{}_{A},\dots  ).
\end{equation}
Obviously, the functions $L^{i}{}_{j}$ above are components 
of the field $L$ with respect to local coordinates $x^{i}$,
\begin{equation}\label{degrees 68}
L=L^{i}{}_{j}(x)\frac{\partial }{\partial x^{i}} \otimes dx^{j}.
\end{equation}
 Obviously, this action is essentially local in the 
sense of $x$-dependence of $L^{i}{}_{j}$ on $x^{k}$. In general
 there is no coordinate system in which they would 
 be constant. And if accidentally they happen to be so in 
some coordinates, this is an artefact; in other coordinates
they will depend on $x^{k}$ (unless $L^{i}{}_{j}$ has the form 
$\lambda \delta ^{i}{}_{j} $, where $\lambda$ is constant).
\newline
The quantities of affine spin, i.e., $\Sigma^{a}{}_{b}$ are 
Hamiltonian generators of (\ref{degrees 66}) in action on internal
 degrees of freedom. So if $L_{x}=\exp (\alpha (x) ) $
$\alpha$ being a mixed tensor field on $M$ then (\ref{degrees 66}), (\ref{degrees 67})
become
\begin{equation}\label{degrees 69}
F_{x}M\ni e=(\dots  ,e_{A},\dots  )\rightarrow  (\dots  ,\exp (\alpha (x) ) e_{A},...),
\end{equation}
i.e., analytically,
\begin{equation}\label{degrees 70}
(...,x^{a},..;...,e^{i}{}_{A},...)\rightarrow  (...,x^{a},...;...,\left( \exp (\alpha (x) )\right)^{i}{}_{j}(x) e^{j}{}_{A},...).
\end{equation}
This means that the phase-space functions $F$ depending
 only on coordinates but not on their conjugate  momenta, i.e., $F(x,e)$
suffer the transformation rule 
\begin{equation}\label{degrees 71}
F\rightarrow \left( \mathfrak{L}[\alpha]\right)F,
\end{equation}
where the differential operator $\mathfrak{L}[\alpha]$ has the 
following form in action on such functions:
\begin{equation}\label{degrees 72}
\mathfrak{L}[\alpha]F=\{\alpha ^{i}{}_{j}(x)\Sigma ^{j}{}_{i},F\}= \alpha^{i}{}_{j}\{\Sigma^{j}{}_{i},F \}.
\end{equation}
Another system of Poisson brackets,
\begin{equation}\label{degrees 73}
\{P_{i},\Sigma^{k}{}_{j} \}=\Sigma^{l}{}_{j}\Gamma^{k}{}_{li} - \Sigma^{k}{}_{l}\Gamma^{l}{}_{ji}
\end{equation}
show that $P_{i}$ are Hamiltonian generators of 
parallel transports along coordinate lines in $M$,  just
 like $\Sigma^{i}{}_{j} $ (as we have just seen) are basic Hamilton 
generators of the system of groups ${\rm GL}(T_{x}M)$.
One can easily show that (and this information
is already contained in the above ones) for any function 
$F$ depending only on the configuration, i.e., on the $FM$-variables
(but not on $P_{i}$ and $P^{A}{}_{i}$) the following holds:
\begin{equation}\label{degrees 74}
\{ \Sigma^{i}{}_{j},F \}= - E^{i}{}_{j}F = - e^{i}{}_{A} \frac{\partial}{\partial e^{j}{}_{A} }F,
\end{equation}
where
\begin{equation} \label{degrees 75}
E^{i}{}_{j} =   e^{i}{}_{K} e^{L}{}_{j} E^{K}{}_{L}= e^{i}{}_{K} \frac{\partial}{\partial e^{j}{}_{K} }\, .
\end{equation}
For (\ref{degrees 74}), (\ref{degrees 75}) no direct analogue holds for functions
 depending on canonical momenta (excepting the flat-space motion).
\newline
Let us write down the co-moving representation of the above brackets, i.e.,
Poisson brackets for quantities $\widehat{P}_{A}$, $\widehat{\Sigma }%
^{A}{}_{B}$:
\begin{equation}\label{degrees 76}
\left\{ \widehat{P}_{A},\widehat{P}_{B}\right\} =\widehat{\Sigma }%
^{K}{}_{L}R^{L}{}_{KAB}-2\widehat{P}{}_{K}S^{L}{}_{AB},
\end{equation}
\begin{equation}\label{degrees 77}
\left\{ \widehat{\Sigma }^{A}{}_{B},\widehat{P}_{C}\right\} =-\widehat{P}{}%
_{B}\delta ^{A}{}_{C},
\end{equation}
\begin{equation}\label{degrees 78}
\left\{ \widehat{\Sigma }^{A}{}_{B},\widehat{\Sigma }^{C}{}_{D}\right\}
=\delta ^{C}{}_{B}\widehat{\Sigma }^{A}{}_{D}{}-\delta ^{A}{}_{D}\widehat{%
\Sigma }^{C}{}_{B},
\end{equation}
\begin{equation}\label{degrees 79}
\left\{ \Sigma ^{i}{}_{j},\widehat{\Sigma }^{A}{}_{B}\right\} =0,
\end{equation}
where the curvature and torsion symbols with capital indices denote the
co-moving components of the corresponding tensors, so they depend not only
on the spatial coordinates $x^{k}$ but also on the along-fibre coordinates $%
e^{i}{}_{A}$:
\begin{equation}\label{degrees 80}
R^{L}{}_{KAB}=e^{L}{}_{i}R^{i}{}_{jab}e^{j}{}_{K}e^{a}{}_{A}e^{b}{}_{B},
\end{equation}
\begin{equation}\label{degrees 81}
S^{K}{}_{AB}=e^{K}{}_{i}S^{i}{}_{jm}e^{j}{}_{A}e^{m}{}_{B}.
\end{equation}
Let us observe both similarities and differences in the structure of the
co-moving system brackets and that one based on (\ref{degrees 62}, \ref{degrees 64}, \ref{degrees 73}). Namely (\ref{degrees 76}, \ref{degrees 77}, \ref{degrees 78})
is very similar to the basic commutation relations for the affine group $%
GAf\left( n,\mathbb{R}\right) $. Indeed, on the right- hand sides of (\ref{degrees 76}, \ref{degrees 77}, \ref{degrees 78}) we
"almost" recognize the structure constants of $
GAf\left( n,\mathbb{R}\right)\simeq {\rm GL}\left( n,\mathbb{R}\right)\underset{\smile}{\times}\mathbb{R}^{n} $ ($\underset{\smile}{\times}$ denotes the semidirect product of ${\rm GL}\left( n,\mathbb{R}\right)$ 
and $\mathbb{R}^{n}$). The proviso "almost" concerns the
first subsystem, i.e., (\ref{degrees 76}). Its right hand sides
 vanish only if the connection $\Gamma$ is flat, i.e.,
if we consider the usual affine-space-problem.
Surprisingly enough, on the right-hand side
the torsion tensor appears, unlike in (\ref{degrees 62}). This
fact is both mechanically and geometrically interesting
in itself, however for us it is not particularly
interesting, because, as a rule, we shall work in
a Riemann manifold, where $\Gamma$ is the usual
Levi-Civita connection, with vanishing torsion.
Nevertheless, the Riemann-Cartan space, where the
torsion is admitted, is also of some interest, first of
all when the linear defects (dislocations) are admitted \cite{Zor1967}.
But the Poisson brackets (\ref{degrees 77}), (\ref{degrees 78}) do not depend
on geometry of $M$ and correspond exactly to the
structure constants of $GAf\left( n, \mathbb{R}\right)$. It is seen that
$\widehat{P}_{A}$ are Hamiltonian generators of parallel transports
along the $A$-th "ribs" of $A$ (roughly speaking).
And similarly $\widehat{\Sigma}^{A}{}_{B} $ are Hamiltonian generators of
(\ref{degrees 11}, \ref{degrees 14}). There are however some structural
differences between the role of $\Sigma^{i}{}_{j} $ and $\widehat{ \Sigma}^{A}{}_{B}$
in this sense. Namely, one can meaningfully say
that $L$ in (\ref{degrees 11}) is constant, it is just the system of scalars in $M$.
\newline
Of course, one can consider "micromaterial transformations"  
where $L$ is local like in gauge theories, i.e., depends on
coordinates $x^{k}$. But nevertheless there exists a finite -
dimensional subgroup, isomorphic with ${\rm GL}(n,\mathbb{R})$ itself 
and given by constant matrix elements $L^{A}{}_{B}$. Unlike
 this, as mentioned, the constancy of $L^{i}{}_{j}$ in (\ref{degrees 67}, \ref{degrees 68}) 
is ill-defined. For the general, $x$-dependent matrices 
$\left[ L^{A}{}_{B}\right]=\left[(\exp{ }\ell )^{A}{}_{B}\right] $ we have, in a full analogy to 
(\ref{degrees 69}, \ref{degrees 70})
\begin{equation}\label{degrees 82}
F_{x}M\ni e=(...,e_{A},...)\rightarrow (...,e_{B}{L(x)}^{B}{}_{A},...),
\end{equation}
\begin{equation}\label{degrees 83}
\left(...,x^{a},...;...,e^{i}{}_{A}(x),...\right)\rightarrow 
\left(...,x^{a},...;...,e^{i}{}_{B}(\exp{ }\ell )^{B}{}_{A},...\right) .
\end{equation}
In analogy to (\ref{degrees 71}), any function of generalized
 coordinates $F(x,e)$ suffers the transformation rule
\begin{equation}\label{degrees 84}
F\rightarrow\left( \exp \Re[\ell ]\right)F ,
\end{equation}
 where again the differential operator $\Re[\ell ]$ acts on
$F$ as follows:
\begin{equation}\label{degrees 85}
\Re[\ell ]F=\{\ell ^{B}{}_{A}(x)\widehat{\Sigma}^{A}{}_{B},F\}= 
\ell ^{B}{}_{A}(x)\{\widehat{\Sigma}^{A}{}_{B},F\}.
\end{equation}
And one can easily show that for any function $F$
depending only on coordinates $x^{i}$, $e^{i}{}_{A}$ we have
\begin{equation}\label{degrees 86}
\{\widehat{\Sigma}^{A}{}_{B},F\}=-E^{A}{}_{B}F=-e^{i}{}_{B}\frac{\partial F }{\partial e^{i}{}_{A} } ,
\end{equation}
\begin{equation}\label{degrees 87}
\{\widehat{P}_{A},F\}=-H_{A}F, \qquad \{P_{i},F\}=H_{i}F ,
\end{equation}
where \cite{Mars1999}
\begin{equation}\label{degrees 88}
H_{i}=e^{A}{}_{i}H_{A}=\frac{\partial }{\partial x^{i}}-\Gamma ^{k}{}_{ji}e^{j}{}_{B}\frac{\partial }{\partial e^{k}{}_{B}} .
\end{equation}
It is seen again that $P_{i}$ are Hamiltonian generators
of parallel transports along the axis of $i$-th coordinates $x^{i}$
and $\widehat{P}_{A} $ are generators of parallel transports along
the $A$-th legs of linear frames $e$.
\newline 
Let us remind the geometric meaning of the
above quantities. In geometrical formulation of the theory
of affine connection the vector fields $E^{K}{}_{L}$ are so-called
fundamental fields of connection acting vertically along
fibres in $FM$. Similarly, $H_{L}$ are basic horizontal
vector fields. Their structure; let us repeat it \cite{Mars1999}:
\begin{equation}\label{degrees 89}
E^{K}{}_{L}=e^{i}{}_{L}\frac{\partial}{\partial e^{i}{}_{K}}, \qquad 
H_{L}=e^{i}{}_{L}\left(\frac{\partial}{\partial x^{i}}-\Gamma^{k}{}_{ij}e^{j}{}_{A}\frac{\partial}{\partial e^{k}{}_{A}}\right)
\end{equation}
implies that their dual system of covector fields,
i.e., differential one-forms $\omega^{K}_{L}$, $\theta^{K}$ is
given by:
\begin{equation} \label{degrees 90}
\omega^{K}{}_{L}=e^{K}{}_{i}\left(de^{i}{}_{L}+\Gamma^{i}{}_{jk}e^{j}{}_{L}dx^{k}\right), \qquad \theta^{K}=e^{K}{}_{i}dx^{i}.
\end{equation}
As usual, by duality we mean the Kronecker-
delta structure of mutual contractions of basic
covariant covectors fields with basic vectors, so:
\begin{equation}\label{degrees 911}
\langle \omega^{K}{}_{L},E^{A}{}_{B}\rangle=\delta^{K}{}_{B}\delta^{A}{}_{L}, \qquad \langle \omega^{K}{}_{L},H_{A}\rangle=0 ,
\end{equation}
\begin{equation}\label{degrees 912}
\langle \theta^{K},E^{A}{}_{B}\rangle=0, \qquad \langle \theta^{K} , H_{A}\rangle=\delta^{K}{}_{A}.
\end{equation}
Let us observe the characteristic geometric duality.
In (\ref{degrees 89}) $E^{K}{}_{L}$ do not depend on the affine connection $\Gamma$. 
They have to do only with the action of ${\rm GL}(n, \mathbb{R})$
along fibres (they manipulate with internal degrees of
freedom). Unlike this, by the very definition of being
horizontal, $H_{K}$ depend explicitly on the affine connection.
And quite dually $\left[\omega^{K}{}_{L} \right]$ defining horizontal fields
as its kernel, is explicitly built of $\Gamma$. But $\theta^{K}$ are
also $\Gamma$-independent; they have to do only with the
bundle structure of internal degrees of freedom. Incidentally
$\theta^{K}$ do not depend on the choice of coordinates, although
apparently they are defined with the use of coordinates
$x^{i}$, but they, as a matter of fact, do not depend on the
choice of $x^{i}$. They may be defined without any use
of $M$-coordinates at all, but here there is no place
for discussing such purely geometric facts \cite{Mars1999}.
Let us observe that the above objects (\ref{degrees 89}), (\ref{degrees 90}) in spite
of their seemingly abstract geometric nature are
nicely interpretable in purely mechanical terms.
Namely, let $\varrho : \mathbb{R}\rightarrow FM$ be a curve describing the
total motion of our object and let $\dot{\varrho}(t)\in T_{\varrho (t)}FM$
be its generalized velocity at the same time instant $t\in \mathbb{R}$.
One can easily show that evaluating the differential
forms $\omega^{K}{}_{L}$, $\theta^{K}$ on $\dot{\varrho}$, i.e., performing the total
contractions of covectors $\omega^{K}{}_{L}$, $\theta^{K}$ with the
vector $\dot{\varrho}(t)$, one obtains just the object of affine
and translational velocities like (\ref{degrees 491}), (\ref{degrees 492}), (\ref{degrees 52}). Indeed,
let us remind that the components of (\ref{degrees 43}) represent
vectors in $FM$ but not in $M$ (except the components
of translational velocity). But evaluating $\omega^{K}{}_{L}$, $\theta^{K}$,
we obtain well-defined co-moving components of
the affine and translational velocity. Indeed,
\begin{equation}\label{degrees 92}
\langle\omega ^{K}{}_{L}, \dot{\varrho}\rangle=\widehat{\Omega}^{K}{}_{L}, \qquad \langle \theta^{K},\dot{\varrho}\rangle=\widehat{v}^{K}.
\end{equation}
Similarly, let us introduce instead (\ref{degrees 89}, \ref{degrees 90})
the corresponding expressions:
\begin{equation}\label{degrees 93}
\begin{array}{c}
E^{i}{}_{j}=e^{i}{}_{K}E^{K}{}_{L}e^{L}{}_{i}=e^{i}{}_{A}\frac{\partial}{\partial e^{j}{}_{A}} ,\\ {}\\
H_{i}=e^{M}{}_{i}H_{M}= \frac{\partial}{\partial x^{i}}-\Gamma^{k}{}_{ji}e^{j}{}_{A}\frac{\partial}{\partial e^{k}{}_{A}} ,\\ {}\\
\omega^{i}{}_{j}=e^{i}{}_{A}\omega ^{A}{}_{B}e^{B}{}_{j}=e^{K}{}_{j} \left( de^{i}{}_{K}+\Gamma^{i}{}_{jm}e^{j}{}_{K}dx^{m}\right) ,\\ {}\\
\theta^{i}=e^{i}{}_{K}\theta^{K}=dx^{i}.
\end{array}
\end{equation}
This is also system of dual bases 
(fields of frames in $FM$),
\begin{equation}\label{degrees 94}
\begin{array}{cc}
\langle\omega^{i}{}_{j}, E^{k}{}_{l}\rangle=\delta^{i}{}_{l}\delta^{k}{}_{j} ,& \langle\omega^{i}{}_{j}, H_{k}\rangle=0,\\ {}&{}\\
\langle\theta^{i}, E^{k}{}_{l}\rangle=0 , &\langle\theta^{i}, H_{k}\rangle=\delta^{i}{}_{k}
\end{array}
\end{equation} 
The difference in comparison with (\ref{degrees 89}, \ref{degrees 90}) is that
(\ref{degrees 94}) are assigned to the particular choice of
coordinates in $M$ and the corresponding induced
coordinates in $FM$, whereas (\ref{degrees 89}, \ref{degrees 90}) are "objective",
i.e., coordinate-independent. But nevertheless, in
analogy to (\ref{degrees 92}),
\begin{equation}\label{degrees 95}
\langle \omega^{i}{}_{j},\dot{\varrho} \rangle=\Omega^{i}{}_{j},\qquad \langle\theta ^{i},\dot{\varrho}\rangle=v^{i}=\frac{dx^{i}}{dt}.
\end{equation}
Roughly speaking (this is a "joke"), one
can write:
\begin{eqnarray}\nonumber
\Omega^{i}{}_{j}=\frac{d\omega^{i}{}_{j}}{dt},\qquad v^{i}=\frac{d\theta^{i}}{dt},
\\\nonumber
\widehat{d\Omega}^{A}{}_{B}=\frac{\omega^{A}{}_{B}}{dt},\qquad \widehat{v}^{A}=\frac{d\theta^{A}}{dt} .
\end{eqnarray}
The doubled $g$-skewsymmetric part of $\Sigma^{i}{}_{j} $ and $\eta $-skewsymmetric
part of $\widehat{\Sigma}^{A}{}_{B}$ are the usual, i.e., metrical, spin quantities.
They are given, by analogy to angular velocities, (\ref{degrees 491},  \ref{degrees 492})
by expressions:
\begin{equation}\label{degrees 96}
S^{i}{}_{j}:=\Sigma^{i}{}_{j}-\Sigma_{j}{}^{i}=\Sigma^{i}{}_{j}-g^{ik}g_{jl} \Sigma^{l}{}_{k},
\end{equation}
\begin{equation}\label{degrees 97}
\widehat{V}^{A}{}_{B}:=\widehat{\Sigma}^{A}{}_{B}-\widehat{\Sigma}_{B}{}^{A} =\widehat{\Sigma}^{A}{}_{B}-\eta^{AC}\eta_{BD}\widehat{\Sigma}^{D}{}_{C}.
\end{equation}
More precisely,$S^{i}{}_{j}$ is the usual spin and $\widehat{V}^{A}{}_{B}$
is what for objects in a flat space was called by Dyson
"vorticity". $S^{i}{}_{j}$ and $\widehat{V}^{A}{}_{B} $ are respectively 
Hamiltonian generators of left acting (spacial) and 
right-acting (micromaterial) rotations preserving respectively
 $g(x)$ and $\eta$. Obviously, we mean the action on internal 
degrees of freedom, and by rotations we meant that 
in (\ref{degrees 66}) $L_{x}$ is $g_{x}$- orthogonal, $L_{x}\in O(T_{x}M,g_{x})$, 
and in (\ref{degrees 11}, \ref{degrees 14}, \ref{degrees 82}) $L$ is $\eta$-orthogonal,
 $L\in O(\mathbb{R}^{n},\eta)$, i.e., analytically,
\begin{equation}\label{degrees 98}
{g_{x}}_{ab}{L_{x}}^{a}{}_{i}{L_{x}}^{b}{}_{j}={g_{x}}_{ij},\qquad \eta _{AB} L^{A}{}_{K}L^{B}{}_{M}=\eta_{KM}.
\end{equation}
It is important to stress that unlike in (\ref{degrees 57}), 
for deformative motion, $\widehat{V}^{A}{}_{B}$ are not co-moving 
components of $S^{i}{}_{j}$:
\begin{equation}\label{degrees 99}
S^{i}{}_{j}\neq e^{i}{}_{A}\widehat{V}^{A}{}_{B}e^{B}{}_{j};
\end{equation}
obviously, the same negative statement is true for
extended bodies in a flat affine space \cite{JJSco2004}, \cite{IJJSco2004}.
\newline
Let us stress a very important point: In flat
affine spaces there exist concepts like the total and
orbital affine momentum (hypermomentum) and total
affine momentum with respect to some fixed spatial
point. The same is true for the usual (metrical) spin
or angular momentum. If $x^{i}$ are coordinates of
the radius-vector of the centre of mass position $x\in M$
with respect to the mentioned fixed origin $\mathfrak{o}\in M $,
\begin{equation} \label{degrees 100}
\overrightarrow{\mathfrak{o}x}=x^{i}e_{i}
\end{equation}
($e_{1},...,e_{n}$ are basic vectors in the translation space $V$ 
of $M$), then the total affine momentum with respect 
to $\mathfrak{o}$ is analytically given by
\begin{equation}\label{degrees 101}
J^{i}{}_{j}=x^{i}P_{j}+\Sigma^{i}{}_{j},
\end{equation}
where obviously, the first term is the "orbital" affine 
momentum. Similarly, for the total angular momentum,
i.e., the doubled skew-symmetric part of $J^{i}{}_{j}$, we 
have the splitting into the "orbital" angular momentum
 $L^{i}{}_{j}$ and spin $S^{i}{}_{j}$, i.e., angular
momentum of the 
body with respect to the instantaneous position of
 the centre of mass in $M$:
\begin{equation}\label{degrees 102}
\mathfrak{J}^{i}{}_{j}=L^{i}{}_{j}+S^{i}{}_{j}.
\end{equation}
As usual, the doubled skew-symmetric parts are meant
in the convention:
\begin{eqnarray}
\nonumber
\mathfrak{J}^{i}{}_{j}&=&J^{i}{}_{j}-J_{j}{}^{i}= J^{i}{}_{j}-g^{ik}g_{jl}J^{l}{}_{k},
\\\label{degrees 103}
L^{i}{}_{j}&=&x^{i}P_{j}-x_{j}P^{i}= x^{i}P_{j}-g^{ik}g_{jl} x^{l}P_{k},
\\
\nonumber
S^{i}{}_{j}&=&\Sigma^{i}{}_{j}-\Sigma_{j}{}^{i}= \Sigma^{i}{}_{j}-g^{ik}g_{jl}\Sigma^{l}{}_{k}.
\end{eqnarray}
In general, only the total quantities are well-balanced,
or in special symmetric cases - just conserved.
Nothing like the splittings (\ref{degrees 101}, \ref{degrees 102}) does exist
in general curved manifolds. Moreover, it is only internal
quantities, i.e., ones related to internal degrees of freedom
that is well-defined. Neither orbital nor total quantities
do exist at all. The reason is that in a general
curved manifold there is no well-defined concept
of the radius-vector.
\newline
Obviously, Green and Cauchy deformation tensor,
$G[e]\in \mathbb{R}^{n*}
  \otimes \mathbb{R}^{n*}$, $C[e]\in{T_{x}}^{*}M \otimes {T_{x}}^{*}M$
(where $e\in F_{x}M$) are given by the usual formulas,
thus, analytically:
\begin{eqnarray}\label{degrees 104}
G[e]_{AB}=g_{ij}(x)e^{i}{}_{A}e^{j}{}_{B},\qquad C[e]_{ij}= \eta_{AB}e^{A}{}_{i}e^{B}{}_{j}.
\end{eqnarray}
These are standard formulas well-known from the 
theory of extended affine bodies in a flat space; 
just the restriction of the general definition to 
homogeneous deformations. However, in both cases 
there is some delicate point the overlooking of which may 
lead to serious mistakes. The trap is hidden in the too 
automatic use of Schouten tensor notation. Namely, 
in certain formulas one uses contravariant tensors 
obtained by the metrical shift of indices, and following 
the Schouten formalism one writes them simply as:
\begin{equation}\label{degrees 105}
G[e]^{AB}=\eta^{AC}\eta^{BD}G[e]_{CD}, \qquad C[e]^{ij}= g^{ik}g^{jl}C[e]_{kl},
\end{equation}
According to the usual convention, when some metrics
$\eta$, $g$ are fixed, the kernel symbols of
tensors with 
$\eta -$ and $g-$ shifted indices are identical with the 
primary ones and it is only the level of the indices that 
encodes the kind of tensor one is dealing with. The 
same convention is used for the inverse metric tensors, 
so, to avoid the crowd of characters one used the symbols
\begin{equation}\label{degrees 106}
\eta_{AB},\quad\eta^{AB} ,\qquad g_{ij},\qquad g^{ij}
\end{equation}
respectively for the covariant metric tensors and their 
contravariant reciprocals. The point is however that in 
certain formulas one uses contravariant tensors reciprocal 
to those with coordinates $G[e]_{AB}$, $C[e]_{ij}$. And 
denoting their components by $G[e]^{AB}$, $C[e]^{ij}$
following (\ref{degrees 106}) would be completely misleading 
because of the confusion with (\ref{degrees 105}). Therefore, for such 
an object one has to introduce a new kernel symbol, 
e.g., $\widetilde{G[e]}$, $\widetilde{C[e]}$ where the following coordinate-independent 
conditions are satisfied:
\begin{equation}\label{degrees 107}
\widetilde{G[e]}^{AC}G[e]_{CB}=\delta ^{A}{}_{B},\qquad \widetilde{C[e]}^{ik}C[e]_{kj}=\delta ^{i}{}_{j}.
\end{equation}
Those conditions are independent of $\eta$,$g$, 
and except some special values of $e$, the inequalities hold: 
\begin{equation}\label{degrees 108}
\widetilde{G[e]}^{AB}\neq G[e]^{AB} ,\qquad \widetilde{C[e]}^{ij}\neq C[e]^{ij}.
\end{equation}
This is one of exceptional situations when the
crowd of characters is unavoidable.
\newline
When postulating and discussing dynamical models
one must use transformation properties of all the
above-introduced quantities under mappings (\ref{degrees 66}),  (\ref{degrees 67}) 
and (\ref{degrees 82}),  (\ref{degrees 11}). It is easy to see that $G[e]$ transforms
under (\ref{degrees 82}), (\ref{degrees 11}) as follows:
\begin{equation}\label{degrees 109}
G[eL(x)]_{AB}=G[e]_{CD}L^{C}{}_{A}(x)L^{D}{}_{B}(x),\quad e\in F_{x}M.
\end{equation}
If for any $x\in M$, $L(x)$ is an orthogonal matrix, 
$L(x)\in O(\mathbb{R}^{n} ,\eta )\subset {\rm GL}(n, \mathbb{R})$ then $C$ is invariant:
\begin{equation}\label{degrees 110}
C[eL(x)]=C[e],\qquad e\in F_{x}M,
\end{equation}
however, there is no particular well-defined rule when
$L(x)$ is a general element of ${\rm GL}(n,\mathbb{R})$.
\newline
Transformation properties under (\ref{degrees 66}),  (\ref{degrees 67}) are dual to the
above ones. So for any mixed tensor field $T$ on $M$ given 
locally by $T^{i}{}_{j}(x)$ $(T(x)\in {\rm GL}(T_{x}M))$, we have 
\begin{equation}\label{degrees 111}
C[T(x)e]_{ij}=C[e]_{kl}T(x)^{k}{}_{i}T(x)^{l}{}_{j},\quad e \in F_{x}M.
\end{equation}
If for any $x\in M$, $T(x)$ is an isometry in the
$g(x)$-sense,  
$T(x)\in O(T_{x}M, g(X))$ then $G[e]$ is invariant:
\begin{equation} \label{degrees 112i}
G[T(x)e]=G[e],\qquad e\in F_{x}M,
\end{equation}
and there is no well-defined rule expressing $G[T(x)e]$
through $G[e]$ alone if $T(x)$ is not an isometry.
 Those are exactly the familiar rules well-known  
from mechanics of extended affine bodies in a flat 
space. Similarly, deformation invariants for internal 
degrees of freedom are given by the classical formulas. 
So, for example, we introduce the mixed Green tensors,
\begin{equation}\label{degrees 112}
\widehat{G}[e]^{A}{}_{B}:=\eta ^{AC}G[e]_{CB}
\end{equation}
and one of possible choice of basic invariants is the
following
\begin{equation}\label{degrees 113}
\mathfrak{K}_{a}[e]:={\rm Tr}\left(\widehat{G}[e]^{a}\right), \qquad a=1,...,n.
\end{equation} 
One can easily show that
\begin{equation} \label{degrees 114}
\mathfrak{K}_{a}[e]:={\rm Tr}\left(\widehat{C}[e]^{-a}\right), \qquad a=1,...,n, 
\end{equation} 
where
\begin{equation}\label{degrees 115}
\widehat{C}[e]^{i}{}_{j}:=g ^{ik}C[e]_{kj}.
\end{equation}
Due to the Cayley-Hamilton theorem, quantities
$\mathfrak{K}_{a}[e]$ constructed according to the rule (\ref{degrees 113}) but with 
other values of $a\in \mathbb{Z}$ may be expressed as functions of
the above ones.
\newline
In spite of the above complete analogy with mechanics
of extended affine bodies in flat space, it must be stressed, however
that in transformation rules of kinematical quantities
some essential changes appear in comparison with the
flat space theory. And this is again a kind of "trap".
Indeed, careful calculations show that (\ref{degrees 66}, \ref{degrees 67})
affect velocities as follows:
\begin{eqnarray}\nonumber
'V^{i}&=&V^{i},
\\ \label{degrees 116}
'\Omega ^{i}{}_{j}&=&T^{i}{}_{l}\Omega^{l}{}_{m}{T^{-1}}^{m}{}_{j}+
V^{k}\left( \nabla_{k}T^{i}{}_{m}\right){T^{-1}}^{m}{}_{j}
\\ \nonumber
&=&T^{i}{}_{l}\Omega^{l}{}_{m}{T^{-1}}^{m}{}_{j}
+ \left( \nabla_{V}T^{i}{}_{m}\right){T^{-1}}^{m}{}_{j},\\\nonumber
'\widehat{\Omega}^{A}{}_{B}&=&\widehat{\Omega}^{A}{}_{B}+e^{A}{}_{l} 
{T^{-1}}^{l}{}_{i}
 \left( \nabla_{V}T^{i}{}_{j}\right)e^{j}{}_{B},
\end{eqnarray}
where the "primed" symbols denote the
$T$-transformed 
quantities and $\nabla_{V}$ is the directional covariant derivative 
along the translational velocity $V$. This rule becomes 
identical with that for flat space only if $T$ is 
covariantly constant, i.e.,
\begin{equation}\label{degrees 117}
\nabla_{k}T^{i}{}_{m}=0.
\end{equation}
The dual canonical momenta transform as follows:
\begin{eqnarray}\nonumber
'P_{i}&=&P_{i}-\Sigma^{k}{}_{l}{T^{-1}}^{l}{}_{j}\nabla_{i}T^{j}{}_{k},
\\\label{degrees 118}
'\Sigma^{i}{}_{j}&=&T^{i}{}_{k}\Sigma^{k}{}_{m}{T^{-1}}^{m}{}_{j},
\\\nonumber
'\widehat{\Sigma}^{A}{}_{B}&=&\widehat{\Sigma}^{A}{}_{B}.
\end{eqnarray}
It is seen that the rule for internal quantities 
(spatial and co-moving components of the affine spin) 
is identical with that for flat spaces. Unlike this, 
the translational covariant momentum suffers an 
additive correction linear in $\nabla T$.
\newline
The micromaterial local transformations (\ref{degrees 82}) act as follows:
\begin{eqnarray}\nonumber
'V^{i}&=&V^{i},
\\\label{degrees 119}
'\Omega ^{i}{}_{j}&=& \Omega ^{i}{}_{j}+e^{i}{}_{B} \left( L^{B}{}_{A,k}{L^{-1}}^{A}{}_{C}\right)e^{C}{}_{j}V^{k},
\\ \nonumber
'\widehat{\Omega}^{A}{}_{B}&=&{L^{-1}}^{A}{}_{C}\widehat{\Omega}^{C}{}_{D} L^{D}{}_{B}  +{L^{-1}}^{A}{}_{C}L^{C}{}_{B,k}V^{k},
\end{eqnarray}
\begin{eqnarray}\nonumber
'P_{i}&=&P_{i}-\widehat{\Sigma}^{A}{}_{C}L^{C}{}_{B,i}{L^{-1}}^{B}{}_{A},
\\ \label{degrees 120}
'\widehat{\Sigma}^{A}{}_{B}&=&{L^{-1}}^{A}{}_{C} \widehat{\Sigma}^{C}{}_{D}L^{D}{}_{B},
\\ \nonumber
'\Sigma^{i}{}_{j}&=&\Sigma^{i}{}_{j},
\end{eqnarray}
where, obviously, comma denotes the usual partial differentiation
of scalar functions. It is seen again that
the affine spin transforms in a "proper"
way, i.e., just
like in a flat space. Translational covariant momentum
and velocity variables transform "properly" when $L$ is
constant, i.e., when we deal with transformations (\ref{degrees 11}).
\bigskip
\section{Equations of motion} 
\bigskip
The first step is to derive equations of motion for 
non-dissipative, Lagrangian-Hamilton systems. Later 
on one introduces the general models by admitting some 
auxiliary generalized forces of non-Hamiltonian nature,
responsible for non-conservative phenomena. The simplest 
way is to use the basic Poisson brackets introduced above. 
Kinetic energy of extended affine bodies in a flat space \cite{JJSco2004}, \cite{IJJSco2004}
suggests us, by the simple analogy, to use the following formula 
\begin{equation}\label{eq 121}
T=T_{tr}+T_{\rm int}= \frac{m}{2}g_{ij}v^{i}v^{j}+
\frac{1}{2}g_{ij}V^{i}{}_{A}V^{j}{}_{B}J^{AB},
\end{equation}
where, let us remind, the positive constant $m>0$ is the 
mass of the body (inertia of translational motion), and $J^{AB}$ 
are components of constant, symmetric ($J^{AB}=J^{BA}$) and 
positively definite micromaterial inertial tensor (inertia of 
the rotational and deformative motion). But when dealing 
with internal degrees of freedom $J$ is a primary quantity, 
no longer the second-order (quadrupole) momentum of the 
mass distribution. Although such an interpretation may be 
admissible from the point of view of the mentioned limit 
transition with the size of the body, there are situations when 
such a procedure is essentially inadequate, e.g., when dealing 
with such objects like gas bubbles in fluids, etc.. They have 
some inertial properties as elementary observation shows, but 
it is hardly expected that the internal inertia in such 
"exotic" situations may be "derivable" from something like 
\begin{equation}\label{eq 122}
J^{AB}=\int a^{A}a^{B}d\mu (a),
\end{equation}
where $a^{K}$ are Lagrange coordinates and the positive 
measure $\mu$ describes the mass distribution. 
It may be interesting and instructive to rewrite (\ref{eq 121}) 
in some modified forms, e.g., 
\begin{eqnarray}
\label{eq 123}
T=T_{{\rm tr}}+T_{{\rm int}}&=&\frac{m}{2}G_{AB}\widehat{v}^{A}\widehat{v}^{B}+
\frac{1}{2}G_{KL}\widehat{\Omega }^{K}{}_{A}\widehat{\Omega }^{L}{}_{B}J^{AB}
\\\nonumber
&=&\frac{m}{2}g_{ij}v^{i}v^{j}+\frac{1}{2}g_{ij}\Omega ^{i}{}_{k}
\Omega ^{j}{}_{l}J[\varphi ]^{kl},
\end{eqnarray}
where $J[\varphi ]^{ij}$ are spatial, thus variable, components of the 
inertial tensor, 
\begin{equation}\label{eq 124}
J[\varphi ]^{kl}= \varphi ^{k}{}_{A} \varphi ^{l}{}_{B} J^{AB} .
\end{equation}
After performing the Legendre transformations for 
Lagrangians of the form $L=T-\mathfrak{U}(x^{i},e^{j}{}_{A})$ (no generalized, 
i.e., velocity-dependent potentials like, e.g., magnetic 
ones) and expressing generalized velocities by canonical 
momenta, we obtain the following expression for the kinetic 
energy (\ref{eq 121}) :
\begin{equation}\label{eq 125}
\mathfrak{T}=\mathfrak{T}_{{\rm tr}} + \mathfrak{T}_{{\rm int}}= \frac{1}{2m}g^{ij}p_{i}p_{j} + 
\frac{1}{2} \widetilde{J}_{AB}P^{A}{}_{i}P^{B}{}_{j}g^{ij},
\end{equation}
where $p_{i}$, $P^{A}{ }_{i}$ are canonical momenta conjugate 
respectively to $x^{i}$, $e^{i}{ }_{A}$ and $\widetilde{ J}$ is the covariant 
inverse of $J$, 
\begin{equation}\label{eq 126}
J^{AC}\widetilde{J}_{CB}=\delta ^{A}{}_{B}.
\end{equation}
{\bf{Warning}}: $\widetilde{J}$ must be not confused with the "covariant 
$\eta$-shift of $J$", i.e., 
\begin{equation}\label{eq 127}
\widetilde{J}_{AB} \neq \eta _{AC} \eta _{BD} J^{CD} .
\end{equation}
More precisely, (\ref{eq 125}) is related to (\ref{eq 121}) by the 
following explicit expression for the Legendre transformation:
 \begin{eqnarray}\label{eq 1281}
p_{i} &=& \frac{\partial L}{\partial \dot{x}^{i}}= \frac{\partial T}{\partial \dot{x}^{i}} = mg_{ij}\frac{dx^{j}}{dt} , \\ \label{eq 1282}
P^{A}{}_{i} &=& \frac{\partial L}{\partial \dot{e}^{i}{}_{A}}= \frac{\partial T}{\partial \dot{e}^{i}{}_{A}} = g_{ij}\frac{d\varphi ^{j}{}_{B}}{dt}J^{BA}.
\end{eqnarray}
Energy is given by the usual formula 
\begin{equation} \label{eq 129}
E= \dot{x}^{i}\frac{\partial L}{\partial \dot{x}^{i}}+
 \dot{e}^{i}{}_{A}\frac{\partial L}{\partial \dot{e}^{i}{}_{A}}-L= T+\mathfrak{U}(x^{i},e^{j}{}_{B})
\end{equation}
and after Legendre transformation it becomes Hamiltonian 
\begin{equation}\label{eq 130}
H = \mathfrak{T}+\mathfrak{U}(x^{i},e^{j}{}_{B}).
\end{equation}
In analogy to (\ref{eq 123}) the kinetic Hamiltonian (\ref{eq 125}) 
may be written in the form 
\begin{eqnarray}\label{eq 131}
\mathfrak{T} = \mathfrak{T}_{tr} + \mathfrak{T}_{\rm int} &=& 
\frac{1}{2m}\widetilde{G}^{AB} \widehat{p}_{A}\widehat{p}_{B} + 
\frac{1}{2} \widetilde{J}_{AB} \widehat{\Sigma}^{A}{}_{C} \widehat{\Sigma}^{B}{}_{D} \widetilde{G}^{CD}\\ \nonumber
&=& \frac{1}{2m}g^{ij} p_{i}p_{j} + 
\frac{1}{2} \widetilde{J[\varphi] }_{kl} \Sigma^{k}{}_{i} 
\Sigma ^{l}{}_{j} g^{ij}.
\end{eqnarray}
This expression is built of geometric quantities-Hamiltonian 
generators of important transformations acting in the 
configuration and phase spaces. More precisely, it is a quadratic 
form of generators. Coefficients of this quadratic form 
depend on the configuration variables $x^{i}$, $e^{i}{}_{A}$, i.e., 
on the position of the body in $M$ and internal parameters. 
\newline
The simplest way to derive equations of motion is not to 
derive them from Lagrange equations. This would be the 
terrible, very strenuous work where mistakes are simply 
unavoidable and the structure of resulting equations is 
non-readable. Rather, at least for non-dissipative systems,
the best procedure is to use the above-quoted basic 
Poisson brackets and to write equations of motion 
in the form 
\begin{equation}\label{eq 132}
\frac{dF}{dt}=\{F, H\},
\end{equation}
where $H$ is the Hamiltonian, e.g., (\ref{eq 130}, \ref{eq 131}) and $F$ 
runs over some complete systems of functionally independent 
functions on the phase space (some non-traditional in the 
sense of Darboux coordination of our phase space). 
Later on, one may perform the inverse Legendre transformation 
and go back to the usual second-order equations of motion
equivalent to Lagrange equations of the second kind. 
For certain reasons it is instructive to begin with 
a rather academic situation when there is no a priori
assumed relationship between affine connection $\Gamma $ and 
metric tensor $g$ in our space manifold $M$. Then 
as a matter of fact we are given two affine connections: 
$\Gamma ^{i}{}_{jk}$ and Levi-Civita connection $\{^{i} _{jk}\}$ induced by $g$.
Obviously, as usual, the difference of two affine connections 
\begin{equation}\label{eq 133}
\mathfrak{K}^{a}{}_{bc}:=\Gamma ^{a}{}_{bc}-\{^{a} _{bc}\}
\end{equation}
is a tensor field (once contravariant, twice covariant) 
although $\Gamma ^{a}{}_{bc}$, $\{^{a}{}_{bc}\}$ separately have not this property.
\newline
Performing the above Poisson-bracket-procedure (\ref{eq 132}) 
of deriving equations of motion and inverting Legendre 
transformation, one obtains the following suggestive system 
of second-order differential equations written in a balance-like 
form:
\begin{eqnarray}\label{eq 134}
m\frac{Dv^{i}}{Dt}&=&m\mathfrak{K}^{i}{}_{jk}v^{j}v^{k}+\Sigma ^{m}{}_{n} R^{n}{}_{m}{}^{i}{}_{j}v^{j} \\
&& \nonumber 
-\mathfrak{K}_{mn}{}^{i}\frac{De^{m}{}_{K}}{Dt}\frac{De^{n}{}_{L}}{Dt}J^{KL}+F^{i},
\end{eqnarray}
\begin{equation}\label{eq 135}
e^{i}{}_{K}\frac{D^{2}e^{j}{}_{L}}{Dt^{2}}J^{KL}=-e^{i}{}_{K}g^{jm}
\frac{Dg_{mn}}{Dt}\frac{De^{n}{}_{L}}{Dt}J^{KL}+N^{ij},
\end{equation}
where the meaning of symbols is as follows:
\begin{enumerate}
\item the shift of indices (up–-down) is meant in the
sense of metric $g$,
\item
\begin{equation}\label{eq 136}
N^{ij}=N^{i}{}_{k}g^{kj}, \qquad N^{i}{}_{k}=-e^{i}{}_{A} \frac{\partial 
\mathfrak{U}}{\partial e^{k}{}_{A}}=-E^{i}{}_{k}\mathfrak{U}
\end{equation} (cf (\ref{degrees 75})),
\item
\begin{equation}\label{eq 137}
F^{i}=g^{ik}F_{k}, \qquad F_{k}=-H_{k} 
\mathfrak{U}=-\frac{\partial \mathfrak{U}}{\partial x^{k}}
+ \Gamma ^{a}{}_{bk}e^{b}{}_{B}\frac{\partial \mathfrak{U}}{\partial e^{a}{}_{B}}
\end{equation} (cf (\ref{degrees 88})).
\end{enumerate}
Let us mention that $H_{k}\mathfrak{U}$ are components of what 
is known in mechanics of principal bundles of frames 
as a covariant exterior differential $D\mathfrak{U}$ of the scalar 
function $\mathfrak{U}$. Note that if $\mathfrak{U}$ is not projectable from 
$FM$ to $M$ (i.e., if it depends not only on $x^{i}$ but 
also on $e^{i}{}_{A}$), then $D\mathfrak{U}\neq d\mathfrak{U}$ unless the affine 
connection $\Gamma$ is flat. Let us notice also that in a 
flat affine space (Euclidean space) our equations of 
motion (\ref{eq 134}), (\ref{eq 135}) reduce to the well-known equations 
of motion of extended affinely rigid bodies \cite{JJSco2004}, \cite{IJJSco2004}. And 
also in curved spaces the main ideas of interpretation 
are similar to the mentioned ones. On the left-hand 
side of (\ref{eq 134}) we have covariant acceleration 
multiplied by mass. Indeed, $v^{i}=\frac{dx^{i}}{dt}$ is just the 
usual velocity of translational motion. On the right-hand 
side we have the usual force $F^{i}$ acting on the 
material point. Similarly, in (\ref{eq 135}) $N^{ij}$ is just 
what was called affine momentum, or hyperforce,
influencing directly the motion of internal degrees of 
freedom. We have started from variational principle, 
therefore, these dynamical quantities $F^{i}$, $N^{ij}$  
have the very peculiar potential structure being build of 
derivatives of $\mathfrak{U}$. Incidentally, let us observe a very 
interesting relationship:
\begin{equation}\label{eq 138}
F_{k}= - \frac{\partial \mathfrak{U}}{\partial x^{k}}- N^{a}{}_{b}\Gamma ^{b}{}_{ak}.
\end{equation}
However, once derived in this way, equations of 
motion (\ref{eq 134}), (\ref{eq 135}) may be easily generalized by 
admitting dissipative phenomena. Simply, we may allow 
the force $F^{k}$ and hyperforce $N^{ab}$ to be completed 
by some completely non-potential terms depending on 
velocities $v^{i}$, $V^{i}{}_{A}$ or equivalently $v^{i}$, $\Omega^{i}{}_{j}$ 
Such additional terms may describe viscous friction, 
both in translational and internal motion. In this 
sense, without the particular model (\ref{eq 136} \ref{eq 137}) 
equations (\ref{eq 134}), (\ref{eq 135}) are quite general.
\newline
Let us now concentrate on peculiarities of motion 
in curved spaces. In a sense the right-hand side 
of (\ref{eq 134}) may be interpreted as th total force $F^{i} _{tot}$ 
affecting translational motion and responsible for 
its covariant acceleration. But except the external 
force $F^{i}$, this total force $F^{i} _{{\rm tot}}$  contains certain 
additional terms. In an obvious way those additional 
terms represent the geometric force in the sense of 
coupling between geometry of $M$ and the particle 
degrees of freedom, both internal and translational one. 
Similarly, in (\ref{eq 135}) besides the usual "external" term 
$N^{ij}$, on the right hand side there is an additional 
geometric term describing the coupling of internal degress 
of freedom with spatial structure. Notice that even 
if external translational force vanishes, $F^{i}=0$, 
translational motion is not geodetic. This is just the 
consequence of the mentioned coupling between geometry 
of $M$ and mechanical degrees of freedom of the 
particle. If we insisted on approximate description 
of a small extended body, this phenomenon would 
have to do with geodetic deviation.
Equations of motion (\ref{eq 134}), (\ref{eq 135}) simplify in a 
remarkable way when we go back to the model 
where affine connection $\Gamma$ is implied by metrical 
structure $g$, or at least when there exists some 
kind of compatibility between $\Gamma$ and $g$. 
\newline
The most natural model, applicable both in defect 
theory \cite{Zor1967} and in General Relativity is that 
of Riemann-Cartan space where metric $g$ is 
assumed to the parallel under affine connection $\Gamma$,
\begin{equation}\label{eq 139}
\nabla_{k}g_{ij}=0.
\end{equation}
Then $\mathfrak{K}^{a}{}_{bc}$ becomes the so-called contorsion 
tensor $K^{a}{}_{bc}$ and it is easily shown in differential 
geometry that \cite{Zor1967}
\begin{equation}\label{eq 140}
K^{a}{}_{bc}=S^{a}{}_{bc}+S_{bc}{}^{a}+S_{cb}{}^{a}
\end{equation}
where $S$ is the torsion tensor of $\Gamma$, 
\begin{equation}\label{eq 141}
S^{i}{}_{jk}=\frac{1}{2} \left(\Gamma ^{i}{}_{jk} - 
\Gamma ^{i}{}_{kj} \right).
\end{equation}
All tensor indices are manipulated in the sense of $g$;
in particular, contorsion is $g$-skew-symmetric in the 
first pair of indices,
\begin{equation}\label{eq 142}
K^{a}{}_{bc}= - K_{b}{}^{a}{}_{c}= - g_{bi} g^{aj}K^{i}{}_{jc}.
\end{equation}
In this way $\Gamma$ is controlled by two independent 
and a priori arbitrary quantities: $g$ and $S$.
Equations of motion (\ref{eq 134}, \ref{eq 135}) simplify then to the 
following form:
\begin{equation}\label{eq 143}
m\frac{Dv^{i}}{Dt}=\Sigma^{a}{}_{b}R^{b}{}_{a}{}^{i}{}_{j}v^{j}+
                   2mv^{b}v^{c}S_{bc}{}^{i}+F^{i},
\end{equation}
\begin{equation}\label{eq 144}
e^{i}{}_{A}\frac{D^{2}}{Dt^{2}}e^{j}{}_{B}J^{AB}=N^{ij}.
\end{equation}
In Riemann-Cartan spaces the curvature tensor is 
g-skew-symmetric in the first pair of indices,
\begin{equation}\label{eq 145}
R^{b}{}_{aij}= - R _{a}{}^{b}{}_{ij}= - 
g_{ac} g^{bd}R^{c}{}_{dij};
\end{equation}
Because of this (\ref{eq 145}) $g$-skewsymmetry it is only 
the $g$-skew-symmetric part of $\Sigma$ that really enters 
the formula (\ref{eq 143}). But this skew-symmetric part is 
just the half of internal angular momentum (spin) in 
the spatial representation, 
\begin{equation}\label{eq 147}
\frac{1}{2}\left(\Sigma ^{i}{}_{j}-\Sigma_{j}{}^{i} \right)
=\frac{1}{2}\left(\Sigma ^{i}{}_{j}-g^{ik}g_{jm}\Sigma^{m}{}_{k} \right)=\frac{1}{2}S^{i}{}_{j}
\end{equation}
cf. (\ref{degrees 96}). Therefore (\ref{eq 143}) becomes:
\begin{equation}\label{eq 148}
\frac{Dp^{i}}{Dt}=\frac{Dv^{i}}{Dt}=2mv^{a}v^{b}S_{ab}{}^{i}+
\frac{1}{2}S^{a}{}_{b}R^{b}{}_{a}{}^{i}{}_{j}v^{j}+F^{i}
\end{equation}
and the similar balance-like form of (\ref{eq 144}) reads 
\begin{equation}\label{eq 149}
\frac{D\Sigma ^{ij}}{Dt}=\widetilde{J}_{ab} \Sigma ^{ai}\Sigma^{bj}+ N^{ij} ,
\end{equation}
where the $e$-dependent inertial tensor $\widetilde{ J}_{ab}$ is given by 
\begin{equation}\label{eq 150}
\widetilde{J}_{ab}=\widetilde{J}_{KL}e^{K}{}_{a}e^{L}{}_{b},
\end{equation}
or, equivalently, in reciprocal contravariant terms
\begin{equation}\label{eq 151}
 J ^{ab}=e^{a}{}_{K}e^{b}{}_{L} J^{KL}.
\end{equation}
Equations of motion (\ref{eq 148}, \ref{eq 149}) are represented here 
 as a system of balance laws for the kinetic linear
momentum and affine spin $p^{i}$, $\Sigma^{ij}$, where according 
to Legendre transformation
\begin{equation}\label{eq 152}
p^{i}=mv^{i},\qquad \Sigma^{i}{}_{j}=g_{ja}\Omega^{a}{}_{b} J^{bi},\qquad 
\Sigma^{ij}=\Omega^{j}{}_{b}J^{bi}.
\end{equation}
Let us stress that unlike in the usual formula 
\begin{equation}\label{eq 153}
P^{A}{}_{i}=g_{ij}\frac{De^{j}{}_{B}}{Dt}J^{BA},
\end{equation}
the "coefficients" $J^{bi}$ in the non--holonomic representation (\ref{eq 152})  depend on the internal configuration $e$. 
So, they are non--constant and are state--dependent 
even in the flat--space theory.
Formally the internal part of dynamics (\ref{eq 144}), (\ref{eq 149}) 
looks like in mechanics of extended affine bodies in 
Euclidean space. The difference is only that the 
usual time derivative $\frac{d}{dt}$ is replaced by the covariant 
one $\frac{D}{Dt}$. The main novelty, as mentioned above,
appears in the dynamics of translational motion 
(\ref{eq 143}), (\ref{eq 148}). Because now it is not only so that 
the "flat" $\frac{dp_{i}}{dt}$ is replaced by $\frac{Dp_{i}}{Dt}$. Namely, in 
addition to the "external" translational force $F^{i}$ two 
"geometric" forces appear on the right--hand side 
of (\ref{eq 143}), (\ref{eq 148}). One of them describes the direct 
coupling between linear momentum (thus also 
translational velocity) and torsion tensor. The other 
one describes the direct coupling between internal 
angular momentum (spin) and the curvature tensor 
$R$. Those geometric forces have the "magnetic--like" 
structure in the sense that they are $g$--orthogonal 
to translational velocity, so they do not do any work. 
The duality: translations--torsion, rotations--curvature 
is well known in differential geometry \cite{Zor1967}, and on the mechanical level--in defect theory (dislocations and 
interstitials or voids) \cite{Bloo1979}, \cite{Zor1967}. This is a nice picture, however, 
let us observe that (\ref{eq 143}, \ref{eq 148}) may be also written 
in shorter form, where the translation–torsion term 
apparently disappears,
\begin{equation}\label{eq 154}
\frac{\delta p^{i}}{\delta t}=m\frac{\delta v^{i}}{\delta t} =
\frac{1}{2}S^{a}{}_{b}R^{b}{}_{a}{}^{i}{}_{j}v^{j}+F^{i},
\end{equation}
where $\frac{\delta}{\delta t}$ denotes the covariant differentiation in the 
$g$--Levi--Civita sense. Obviously, if there is no torsion then 
 $\frac{\delta}{\delta t}=\frac{D}{D t}$.
\newline
The above equations of motion are structurally similar 
to generally--relativistic equations for the pole--dipole 
particle. The mentioned generally relativistic models 
were investigated by Mathisson, Weyssenhoft, Papapetrou,
 Tulczyjew, K\"{u}nzle and many others \cite{51Math}, \cite{Papa1951}, \cite{Tul1962}. The 
characteristic couplings: translations--torsion, rotations--curvature have a very deep geometric background 
and in this way, "geometric forces" on the right--hand 
 side of (\ref{eq 143}, \ref{eq 148}), although derived from some 
dynamical Lagrangian postulates, are based on some 
kind of a priori.
\newline
We concentrated here on the Riemann spaces,  when $\nabla g=0$
and torsion tensor does vanish. Riemann--Cartan 
space is in a sense more natural, although there appear some 
ambiguities in the definition of angular velocity and 
affine velocity when torsion is admitted. We do not 
deal here with more general relationships between 
$\Gamma$ and $g$, although some of them should be mentioned 
as interesting in some possible future investigations.
 Riemann--Cartan--Weyl space is defined as such one 
in which parallel transport preserves angels between 
vectors but not necessarily their lengths (roughly speaking 
parallel transport acts on vectors as a conformal transformation). 
Then we have 
\begin{equation}\label{eq 155}
\nabla_{k}g_{ij}=-Q_{k}g_{ij},
\end{equation}
where
\begin{equation}\label{eq 156}
Q^{k}=g^{km}Q_{m}
\end{equation}
is traditionally referred to as the Weyl vector ($Q_{k}$ is then 
the Weyl co-vector). Then (\ref{eq 133}) has the form 
\begin{equation}\label{eq 157}
\mathfrak{K}^{i}{}_{jk}=S^{i}{}_{jk}+S_{jk}{}^{i}+S^{k}{}_{ji}+
\left( \delta ^{i}{}_{j} Q_{k}+\delta ^{i}{}_{k} Q_{j}-g_{jk}Q^{i} \right),
\end{equation}
where, as usual, tensor indices are shifted in the $g$-sense. 
If $\Gamma$ is symmetric (no torsion) then obviously: 
\begin{equation}\label{eq 158}
\mathfrak{K}^{i}{}_{jk}= \delta ^{i}{}_{j} Q_{k}+\delta ^{i}{}_{k} Q_{j}-g_{jk}Q^{i}
\end{equation}
and one is dealing with what is usually called the Weyl 
space. Riemann-Cartan- Weyl spaces and Weyl spaces 
are interesting in themselves and useful in defect 
dynamics \cite{Zor1967}. However, the problem of how 
to define angular and affine velocity becomes here 
essential. In any case in this paper we concentrate 
on Riemann-, or sometimes-  Riemann-Cartan spaces.
In mechanics of extended affine bodies in flat 
spaces one often discusses  the problem of affine 
dynamics with additional constraints \cite{JJSco2004}, \cite{IJJSco2004}, \cite{Woz1988}. The most 
important example is that of gyroscopic motion, 
when the body is metrically-rigid. By analogy we 
can, and in many problems just should, discuss the 
motion of infinitesimal gyroscope in a curved manifold. 
According to gyroscopic constraints the frame $e$ should 
be $g$-orthonormal during the motion; analytically:
\begin{equation}\label{eq 159}
\eta _{AB}e^{A}{}_{i}(t)e^{B}{}_{j}(t)=g_{ij}\left(x(t)\right).
\end{equation}
This means that affine velocity $\Omega ^{i}{}_{j}$ is permanently
$g$-skew-symmetric:
\begin{equation}
\label{eq 160}
\Omega^{i}{}_{j}=-\Omega_{j}{}^{i}=-g_{jk}\Omega^{k}{}_{m}g^{mi},
\end{equation}
just angular velocity in spatial representation. 
The corresponding equations of motion may be obtained  
from (\ref{eq 143} \ref{eq 144}), i.e., from (\ref{eq 148}, \ref{eq 149}), even without the 
use of variational principle, just basing on the 
d'Alambert principle of ideal constraints. 
The power of generalized forces $F^{i}$, $N^{ij}$ on 
 virtual generalized velocities $v^{i}$, $\Omega ^{i}{}_{j}$ is given by 
\begin{equation}\label{eq 161}
\mathfrak{P}=F_{i}v^{i}+N^{i}{}_{j}\Omega^{j}{}_{i}.
\end{equation}
According to the d'Alambert principle, equations of 
gyroscopically  constrained motion are obtained in the 
following way:
\begin{enumerate}
\item
on the right-hand side of unconstrained 
equations one introduces additional reaction forces 
$F_{R}$, $N_{R}$ which maintain constraints, 
\item 
one substitutes formally constraints equations,
\item 
one assumes that $F_{R}$, $N_{R}$ do not do any 
work on virtual displacements compatibile with 
constraints:
\begin{equation}\label{eq 162}
\mathfrak{P}_{R}={F_{R}}_{i}v^{i}+ {N_{R}}^{i}{}_{i}\Omega ^{j}{}_{i}=0
\end{equation}
for any $v$ and any $\Omega$ satisfying (\ref{eq 160}). 
\end{enumerate}
The result is that $F_{R}$ does vanish and $N_{R}$ 
is $g$-symmetric, i.e., its $g$-skew-symmetric part 
does vanish:
\begin{equation}\label{eq 163}
{F_{R}}^{i}=0,\qquad{N_{R}}^{i}{}_{j}-{N_{R}}_{j}{}^{i}=
{N_{R}}^{i}{}_{j}-g_{jk}{N_{R}}^{k}{}_{m}g^{ml}=0
\end{equation}
or briefly:
\begin{equation}\label{eq 164}
{F_{R}}^{i}=0,\qquad{N_{R}}^{ij}-{N_{R}}^{ji}=0.
\end{equation}
This means that the effective reactions-free system 
of equations of motion consist of (\ref{eq 143}) (equivalently 
(\ref{eq 148}, \ref{eq 154}), the skew-symmetric part of (\ref{eq 144}) 
(equivalently – one of (\ref{eq 149})) and constraints 
equations (\ref{eq 159}). The mentioned skew-symmetric 
part of internal equation
\begin{equation}\label{eq 165}
\left( e^{i}{}_{A}\frac{D^{2}}{Dt^{2}}e^{j}{}_{B} -
e^{j}{}_{A}\frac{D^{2}}{Dt^{2}}e^{i}{}_{B}\right)J^{AB}=N^{ij}-N^{ji}
\end{equation}
has the following balance form given by the skew-symmetric part of (\ref{eq 149}) 
\begin{equation}\label{eq 166}
\frac{DS^{ij}}{Dt}=N^{ij}-N^{ji}
\end{equation}
where, obviously, $S^{i}{}_{j}$ are spin components 
(\ref{degrees 96}, \ref{eq 147}). 
The skew-symmetric internal hyperforce 
\begin{equation}\label{eq 167}
\mathfrak{N}^{ij} =N^{ij}-N^{ji}
\end{equation}
is just the usual torque, i.e., moment of forces 
acting on the body. In $n$ dimensions it is just 
a skew–symmetric tensor;  the peculiarity of the physical 
dimension $n$=3 is that it may be identified in a 
known way with the axial vector $\mathfrak{N}^{i}$. 
\newline
Just like in mechanics of extended affine bodies in 
flat spaces one can consider also other physically 
interesting constraints like incompressible motion, 
rotation -less motion etc. The corresponding equations of 
motion consist of (\ref{eq 148}) i.e. (\ref{eq 154}) and respectively the 
trace-less part of symmetric part etc. of (\ref{eq 144}) i.e. (\ref{eq 149}), 
and, of course, equations of constraints themselves.
\newline
The above balance form of equations of motion is 
very instructive and reveals geometric foundations of the 
model, first of all its symmetry properties and the 
corresponding conservation laws. Another problem is how 
to solve equations of motion and determine the phase 
portrait, at least qualitatively. And here the problem of 
using the non-holonomic reference frame $E$ is crucial. 
There are two main reasons for that:
\begin{enumerate}
\item As already mentioned, when gyroscopic or other
constraints are imposed, the quantities $(x^{i},e^{i}{}_{A})$,
or equivalently $(x^{i},e^{A}{}_{i})$ are no longer independent 
generalized coordinates. Then the  best way to introduce 
reasonable and computationally effective generalized 
coordinates is just to introduce an auxiliary field 
of (co)frames $E$, use representation (\ref{degrees 32}), confine 
matrices $L$ to be elements of $SO(n,\mathbb{R})$ (gyroscopic 
model) or some other subgroup $G\subset {\rm GL}(n.\mathbb{R})$, and 
parametrize $G$ with the use of some natural coordinates
e.g., canonical coordinates of the first or second kind, 
Euler angels, some byproducts etc.
\item Even if no additional constraints are imposed on 
affine motion of internal degrees of freedom, it is rather 
a rule than exception that the natural coordinates 
$(x^{i},e^{i}{}_{A})$ on $FM$  are inconvenient and non-effective 
in study of realistic problems like elastic vibrations 
and their coupling with rigid rotations. For example, 
one deals often with isotropic problems when the 
potentials energy $\mathfrak{U}$ is built of deformation invariants. 
The only reasonable procedure is then to use (\ref{degrees 32}) 
and express $L$ in terms of left or right polar 
decomposition or two-polar decomposition (singular 
value decomposition). Incidentally, this enables one 
to use description as similar as possible to the 
one of extended bodies in flat space.
\end{enumerate}
For any  $L\in {\rm GL}(n,\mathbb{R})$ we have two version of the 
polar decomposition, sometimes referred to as the 
left or right one:
\begin{equation}\label{eq 168}
L=OS=\Sigma O,\qquad \Sigma=OSO^{-1},
\end{equation}
where $O\in O(n,\mathbb{R})$ is orthogonal and $S=S^{T}$,  
$\Sigma = \Sigma ^{T}$ are symmetric and positively- definite. 
Performing diagonalization of $S$ with the help 
of some orthogonal matrix $V\in O(n,\mathbb{R})$,
\begin{equation}\label{eq 169}
S=VDV^{-1},
\end{equation}
$D$ being diagonal and positive, and denoting:
\begin{equation}\label{eq 170}
U=OV\in O(n,\mathbb{R})
\end{equation}
we obtain the singular value decomposition 
(two-polar decomposition):
\begin{equation}\label{eq 171}
L=UDV^{-1},\qquad   U,\quad V \in O(n,\mathbb{R}), \qquad D-\mathrm{diagonal}.
\end{equation}
It is well known that the polar decomposition 
is unique, but the singular value decomposition 
suffers some kind of multivaluedness which is 
essentially harmless if properly treated \cite{JJSco2004}, \cite{IJJSco2004}. 
\newline
In (\ref{eq 168}) internal degrees of freedom are 
represented as consisting of two subsystems:
rigid body in $n$-dimensions and deformations; 
respectively $\frac{1}{2}n(n-1)$ and $\frac{1}{2}n(n+1)$  degrees of freedom. 
There are two possible representations of deformative 
modes, as seen in (\ref{eq 168}). The symmetric-deformative 
objects have respectively to do with the Green and 
Cauchy deformation tensors. Namely, matrices of those 
tensors are given by
\begin{equation}\label{eq 172}
G=L^{T}L=S^{2} ,\qquad C=\Sigma^{-2},\qquad C^{-1}=\Sigma ^{2} .
\end{equation}
In the two-polar decomposition $L$ consists  of two 
fictitious rigid bodies represented respectively by $U$ and $V$;
every with $\frac{1}{2}n(n-1)$ degrees of freedom, and of $n$ 
purely scalar deformations (stretching) which tell 
us only how the body is stretched, but without any 
information about orientations of this stretching 
both in physical space and in material of the body.
The quantities $U$, $V$ describe orientations of stretching 
i.e., material and spatial position of the main
axes of Green and Cauchy deformation tensors,
\begin{equation}\label{eq 173}
G=VD^{2}V^{-1},\qquad C=UD^{2}U^{-1}, \qquad C^{-1}=UD^{-2}U^{-1}.
\end{equation}
Having rigid bodies we can introduce their angular 
velocities, both "spatial" and "co-moving" versions. 
Obviously the spatial representations are given by :
\begin{equation}\label{eq 174}
\omega_{{\rm rl}}=\frac{dO}{dt}O^{-1},\qquad
\chi_{{\rm rl}}=\frac{dU}{dt}U^{-1},\qquad
\vartheta_{{\rm rl}}=\frac{dV}{dt}V^{-1},
\end{equation}
respectively for the polar and two-polar decomposition. 
Similarly, the "co-moving" expressions have the known form:
\begin{equation}\label{eq 175}
\widehat{ \omega}_{{\rm rl}}=O^{-1}\frac{dO}{dt},\qquad
\widehat{\chi}_{{\rm rl}}=U^{-1}\frac{dU}{dt},\qquad
\widehat{\vartheta}_{{\rm rl}}=V^{-1}\frac{dV}{dt}.
\end{equation}
Obviously, they are related to each other in the 
usual way, justifying the terms spatial and co-moving 
\begin{equation}\label{eq 176}
\omega _{{\rm rl}}=O\widehat{\omega}_{{\rm rl}} O^{-1},\qquad
\chi _{{\rm rl}}=U\widehat{\chi}_{{\rm rl}} U^{-1},\qquad
\vartheta _{{\rm rl}}=V\widehat{\vartheta}_{{\rm rl}} V^{-1};
\end{equation}
We must remember however that this is something 
a bit else than the usual relationship between, 
e.g., $\Omega$ and $\widehat{\Omega}$.
\newline
The labels $"rl"$ or $"dr"$ at those angular velocities 
refer to the fact that those quantities describe the 
$L$-motion relative with respect to the reference frame 
$E$, usually non-holonomic one. 
It is not only geometrically interesting but also 
computationally effective to express the kinetic energy 
of internal motion $T_{\rm int}$, cf (\ref{eq 121}) through the above 
quantities.
One obtains:
\begin{equation}\label{eq 177}
{T_{\rm int}=-\frac{1}{2}{\rm Tr}\left( SJS\widehat{\omega }^{2}\right) 
+{\rm Tr}\left( SJ\frac{dS}{dt}\widehat{\omega }\right) 
+\frac{1}{2} {\rm Tr}\left( J\left( \frac{dS}{dt}\right) ^{2}\right) },
\end{equation}
where
\begin{equation}\label{eq 178}
\widehat{\omega }=\widehat{\omega }_{{\rm dr}}+\widehat{\omega }_{{\rm rl}}=\widehat{\omega }_{{\rm dr}}+O^{-1}\frac{dO}{dt};
\end{equation}
it is clear that, $\widehat{\omega }_{{\rm dr}}$ is the restriction of $\widehat{\Omega }_{{\rm dr}}$ 
(\ref{degrees 51}) 
to the rigid motion of the $O$-gyroscope,
\begin{equation}
{\widehat{\omega}_{{\rm dr}}}^{A}{}_{B}={O^{-1}}^{A}{}_{F}\Gamma^{F}{}_{DC}U^{D}{}_{B}
U^{C}{}_{E}{\widehat{V}}^{E}.
\end{equation}
{\bf Remark:} do not confuse completely different things 
denoted by the same kernel symbol S: torsion $S^{i}{}_{jk}$,
spin $S^{i}{}_{j}$, and deformation-–symmetric part of 
the polar decomposition $S^{AB}$; unfortunately letters are 
missing.
\newline
In expression (\ref{eq 177}) we have used the standard 
orthonormal coordinates in $\mathbb{R}^{n}$. Then  the micromaterial 
metric is analytically given by the Kronecker symbol,
\begin{equation}\label{eq 180}
\eta_{AB}=\delta_{AB}.
\end{equation}
Obviously, if for any reasons convenient, we can use 
general rectilinear coordinates in $\mathbb{R}^{n}$ (no (\ref{eq 180})). Then 
in (\ref{eq 177}) instead $J$ we must use $J_{\eta}$ given by 
\begin{equation}
{J_{\eta}}^{A}{}_{B}=J^{AC}\eta _{CB}.
\end{equation}
The polar decomposition is convenient when $J$ is 
general and the internal potential energy $\mathfrak{U}$  
is spatially isotropic,
\begin{equation}\label{eq 182}
\mathfrak{U}\left( L(x)  e(x)\right)
= \mathfrak{U}\left(e(x)\right)
\end{equation}
for any isometry $L(x)\in O(T_{x}M, g_{x})$ at any $x\in M$
acting according to (\ref{degrees 67} \ref{degrees 66}). In other words, 
$\mathfrak{U}$ depends on e through the Green deformation 
tensor $G$ (\ref{degrees 104}). 
\newline
The singular value decomposition (two-polar 
decomposition (\ref{eq 171}))is convenient in doubly isotropic 
problems, i.e., ones isotropic both in the physical 
and the micromaterial space. This means two things: 
\begin{enumerate}
\item 
Inertial tensor $J$ is invariant under $O(n,\eta )$
acting through (\ref{degrees 12}, \ref{degrees 14}), therefore,
\begin{equation}\label{eq 183}
J^{AB}=I\eta^{AB}=^{*} I\delta^{AB};
\end{equation}
the last expression based on the natural choice 
of orthogonal coordinates in $\mathbb{R}^{n}$, 
\item 
Potential energy satisfies both (\ref{eq 182}) and 
\begin{equation}
\label{eq 184}
\mathfrak{U}\left(  e(x)L \right)
= \mathfrak{U}\left(e(x)\right)
\end{equation}
for any $L\in O(n,\mathbb{R})$  acting through (\ref{degrees 12}, 
\ref{degrees 14}).
This means that $\mathfrak{U}$ depends on $e$ only through 
deformation invariants (\ref{degrees 113}, \ref{degrees 114}).
\end{enumerate}
\medskip
If those conditions are satisfied, the singular 
value decomposition provides the most effective 
coordinatization of  $FM$ and the formula for kinetic 
energy becomes:
\begin{equation}
\label{eq 185}
T_{\rm int}= - \frac{I}{2}{\rm Tr}\left( D^{2}\widehat{\chi}^{2}\right) -
\frac{I}{2}{\rm Tr}\left( D^{2}\widehat{\vartheta}^{2}\right)+
I{\rm Tr}\left( D\widehat{\chi}D\widehat{\vartheta} \right)+ 
\frac{I}{2} {\rm Tr} \left( \left(\frac{dD}{dt}\right)^{2}\right) ,
\end{equation}
where now:
\begin{eqnarray}\nonumber
\widehat{\vartheta}&=&V^{-1}\frac{dV}{dt},
\\
\label{eq 186}
\widehat{\chi}&=&\widehat{\chi}_{{\rm dr}}+\widehat{\chi}_{{\rm rl}}=
\widehat{\chi}_{{\rm dr}}+U^{-1}\frac{dU}{dt},
\\ \nonumber
\widehat{\chi}_{{\rm dr}}&=& {U^{-1}}^{A}{}_{F}\Gamma ^{F}{}_{DC}L^{D}{}_{B} 
L^{C}{}_{E}\widehat{V}^{E} .
\end{eqnarray}
There is no drive term in $\widehat{\vartheta}$.
Expressions (\ref{eq 177}, \ref{eq 185}) have some very peculiar 
features, interesting from the geometric and analytic 
point of view, and at the same time very convenient 
in physical calculations. Namely, formally they are 
identical with the corresponding formulas for affine 
motion in flat spaces. The difference is that $\widehat{\chi}$ and $\widehat{\omega}$ 
contain the additional "drive" terms $\widehat{\chi}_{{\rm dr}}$, $\widehat{\omega}_{{\rm dr}}$. These 
terms depend on geometry of $M$. Translational velocity 
 occurs in these terms, because of this, they interfere 
somehow in (\ref{eq 121}) with the translational term $T_{tr}$.
\newline
We shall discuss some special examples in 
two-dimensional spaces $n=2$. The peculiarity of dimension 
two is that the group $O(2,\mathbb{R})$ is Abelian-the exception
among the groups $O(n,\mathbb{R})$ because for any $n>2$ they are 
semi-simple. Because of this commutativity, the special and 
co moving representations of angular velocity coincide:
\begin{equation}\nonumber
\widehat{\vartheta}=\vartheta, \qquad 
\widehat{\chi}_{{\rm rl}}=\chi_{{\rm rl}}, \qquad\widehat{\chi}=\chi .
\end{equation}
Because of this, another peculiarity appears, namely, 
there exists an interesting class of integrable models 
with directly separable Hamilton-Jacobi equations.
\newline
However, before doing this, for completeness, we 
quote the co-moving form of balance equations 
(\ref{eq 143}, \ref{eq 144}, \ref{eq 148}, \ref{eq 149}). They are curved-space 
affine counterparts of Euler equations known form 
rigid body mechanics in Euclidean space. Namely, after 
some calculations one can obtain:
\begin{eqnarray}
\label{eq 187}
m\frac{dv^{A}}{dt}&=&-m\widehat{\Omega}^{A}{}_{B}v^{B}+
2v^{B}v^{C}S_{CB}{}^{A}+
\frac{1}{2}S^{C}{}_{D}R^{D}{}_{C}{}^{A}{}_{B}v^{B}+
F^{A},\\
\label{eq 188}
\frac{\widehat{\Omega}^{B}{}_{C}}{dt} J^{CA}
&=&-m\widehat{\Omega}^{B}{}_{D}
\widehat{\Omega}^{D}{}_{C}J^{CA}+N^{AB},
\end{eqnarray}
where the capitals refer to the co-moving representation,
e.g.,$v^{A}=e^{A}{}_{i}v^{i}$,etc., and their raising and lowering 
is meant in the sense of micromaterial metric $\eta_{AB}$
(usually $\delta_{AB}$; rectilinear orthogonal coordinates in $\mathbb{R}^{n}$ are 
most convenient).
\newline
Expressing velocities and affine velocities in terms 
of linear momentum and spin in co-moving 
representation, we obtain the following balance laws:
\begin{eqnarray}
\label{eq 189}
\frac{dp^{A}}{dt}&=&-p^{B}\widetilde{J}_{BC}\widehat{\Sigma}^{CA}+
\frac{2}{m}p^{B}p^{C}S_{CB}{}^{A}+
\frac{1}{2m}S^{D}{}_{C}R^{C}{}_{D}{}^{A}{}_{B}p^{B}+
F^{A},\\
\label{eq 190}
\frac{\widehat{\Sigma}^{AB}}{dt}&=&-\widehat{\Sigma}^{AC}\widetilde{J}_{CD}
\widehat{\Sigma}^{DB}+N^{AB},
\end{eqnarray}
The Euler--like structure of those equations is easily 
readable.
\newline
Obviously, there is nothing wrong in that on the left-
hand side of equations we have the usual differentiation. 
Though from the point of view of geometry of $M$, 
the capital-–indices-–quantities, i.e., co--moving components, 
are just scalars. Their covariant derivatives along curves 
are therefore identical with the usual derivatives.
\bigskip
\section{Examples. Two--dimensional homogeneously deformable body. }
\bigskip
Let us present as interesting examples the two-dimensional
body moving in constant-curvature spaces, i.e., the
spherical space $S^{2}(0,R)$ and pseudo-spherical Lobachevsky space 
$H^{2,2,+} (0,R)$. There are realistic situations when the model is rigorously  solvable. It is a good illustration of how our method of nonholonomic frames works practically. We also obtain an interesting class of models integrable in in the Liouville sense. In "polar" coordinates $(r,\varphi )$ the corresponding
metric elements are given respectively by 
\begin{eqnarray}\label{two 200}
ds^{2}&=& dr^{2} + R^{2}\, \sin ^{2}\left(\frac {r}{R}\right) d\varphi ^{2}, \\ \nonumber
ds^{2} &=& dr^{2} + R^{2}\, \sinh ^{2}\left(\frac {r}{R}\right) d\varphi ^{2}.
\end{eqnarray}
In the spherical case all situations with $r=0$, $r = \pi R$
and arbitrary values of $\varphi $ correspond to the same points, the "north" pole if $r = 0$ and the "south" pole if $r=2\pi$. The range of $r$
is $[0,\pi R]$. 
For the pseudo-spherical Lobachevsky space $H^{2,2,+} (0,R)$
the range of $r$ is $[0,\infty ]$.
The most convenient choice of the auxiliary reference frame is:
\begin{eqnarray}
E_{(r)} = \frac {\partial }{\partial r}, \qquad E_{(\varphi )} = \frac
{1}{R\,\sin\,(\frac {r}{R})} \, \frac {\partial }{\partial \varphi }, \\ E_{(r)} = \frac {\partial }{\partial r}, \qquad
E_{(\varphi )} = \frac {1}{R\,\sinh\,(\frac {r}{R})} \, \frac
{\partial }{\partial \varphi },  \label{two 201}
\end{eqnarray}
respectively, in the spherical and pseudospherical case. These frames are evidently non-holonomic.
The kinetic energy may be written in the form:
\begin{equation}\label{two 202}
T=\frac{m}{2}G_{ij}\dot{q}
^{i}\dot{q}^{j}, 
\end{equation}  
where $ q^{i}$ are six generalized coordinates. We introduce them below; obviously, $(\varphi ,r )$ will be two of them.
 Now we can introduce the canonical formalism: $H=T+V$,
where $T=\frac{1}{2m}G^{ij}p_{i}p_{j}$. The matrix $G^{jk}$ is reciprocal to $\
G_{ij}$, i.e. $\ G_{ij}G^{jk}=\delta _{i}{}^{k}$.
Translational kinetic energies have the form:
\begin{eqnarray}\label{two 203}
\mathrm{sphere:}\qquad T_{tr}&=&\frac{m}{2}\left(  \left(  \frac{dr}{dt}\right)
^{2}+R^{2}\sin ^{2}\left(  \frac{r}{R}\right)  \left(  \frac{d\varphi}
{dt}\right)  ^{2}\right) ,\\ \label{two 204}
\mathrm{pseudosphere:}\qquad T_{tr}&=&\frac{m}{2}\left(  \left(  \frac{dr}{dt}\right)
^{2}+R^{2}\sinh{}^{2}\left(  \frac{r}{R}\right)  \left(  \frac{d\varphi}
{dt}\right)  ^{2}\right).
\end{eqnarray}
When the body is materially isotropic, in the two-dimensional case the inertia tensor has only one essential component $J_{1}=J_{2}=J$. It is convenient to
use the two-polar decomposition $\varphi=UDV^{-1}$, where $U$, $V$ are orthogonal and $D$--is diagonal. Generalized internal
coordinates are $\alpha,\beta,\lambda,\mu,$:
\[
U =\left[
\begin{array}
[c]{cc}
\cos\left(  \alpha\right)  & -\sin\left(  \alpha\right) \\
\sin\left(  \alpha\right)  & \cos\left(  \alpha\right)
\end{array}
\right],\qquad
D=\left[
\begin{array}
[c]{cc}
\lambda & 0\\0 & \mu
\end{array}
\right],\]
\[
 V=\left[
\begin{array}
[c]{cc}
\cos\left(  \beta\right)  & -\sin\left(  \beta\right) \\
\sin\left(  \beta\right)  & \cos\left(  \beta\right)
\end{array}
\right].\]
Then $T_{\rm int}$\ is given by:
\begin{eqnarray}
\label{two 205}
T_{\rm int}&=&\frac{J}{2}\left[  \left(
\frac{d\lambda}{dt}\right)^{2}+\left( \frac{d\mu}{dt}\right)
^{2}\right]  +\frac{J}{2}\left[ \lambda^{2}+\mu^{2}\right]
\chi^{2} \\ \nonumber &+&\frac{J}{2}\left[  \lambda^{2}+\mu^{2}\right]
\vartheta^{2}+2J\lambda \mu\chi\vartheta ,
\end{eqnarray} 
where, $\vartheta=\frac{d\beta}{dt}$\ and $\chi$\ is given by
\begin{equation}\label{two 206}
\chi=\frac{d\alpha}{dt}+\cos\left(  \frac{r}{R}\right)
\frac{d\varphi}{dt}
\end{equation}
on the 2-dimensional sphere and 
\begin{equation}\label{two 207}
\chi=\frac{d\alpha}{dt}+\cosh\left(\frac{r}{R}\right)
\frac{d\varphi}{dt}
\end{equation}
on the 2-dimensional pseudosphere.
It is convenient to introduce new variables $\gamma:=\alpha+\beta$, $\delta:=\alpha-\beta$,
$x:=\frac{1}{\sqrt{2}}\left(  \lambda-\mu\right)$, $y:=\frac{1}{\sqrt{2}}\left(
\lambda+\mu\right)$. Then in the spherical case $T=\frac{m}{2}G_{ij}\frac{dq^{i}}
{dt}\frac{dq^{j}}{dt}$, where $\left[\dots q^{i}\dots \right]  =\left[  r,\varphi,\gamma,
\delta,x,y\right]$\ and the matrix $\left[  G_{ij}\right]$\ is given by
\begin{equation} \label{two 208}
\left[  G_{ij}\right]  =\left[
\begin{array}
[c]{cccc}%
1 & 0 & 0 & 0\\
0 & \left[  \tilde{G}_{kl}\right]  & 0 & 0\\
0 & 0 & \frac{J}{m} & 0\\
0 & 0 & 0 & \frac{J}{m}
\end{array}
\right],
\end{equation}
and  the $3\times 3 $ block $\left[  \tilde{G}_{kl}\right]$ is given by
\[
\left[  \tilde{G}_{kl}\right]  =\left[
\begin{array}
[c]{ccc}
R^{2}\sin^{2}\left(  \frac{r}{R}\right)  +\frac{J\cos^{2}\left(  \frac{r}
{R}\right)  \left(  x^{2}+y^{2}\right)  }{m} & \frac{Jx^{2}\cos\left(
\frac{r}{R}\right)  }{m} & \frac{Jy^{2}\cos\left(  \frac{r}{R}\right)  }{m}\\{}&{}&{}\\
\frac{Jx^{2}\cos\left(  \frac{r}{R}\right)  }{m} & \frac{Jx^{2}}{m} & 0\\{}&{}&{}\\
\frac{Jy^{2}\cos\left(  \frac{r}{R}\right)  }{m} & 0 & \frac{Jy^{2}}{m}
\end{array}
\right].
\]
In the canonical formalism: $T=\frac{1}{2m}G^{ij}p_{i}p_{j}$, $H=T-V$, where
$G_{ij}G^{jk}=\delta_{i}{}^{k}$. One can show that if generalized coordinates are
ordered as $\left[r,\varphi,\gamma,\delta,x,y\right]$, then the matrix $\left[G^{ab}\right]$\
is given by
\begin{equation} \label{two 209}
\left[  G^{ab}\right]  =\left[
\begin{array}
[c]{cccc}
1 & 0 & 0 & 0\\
0 & \left[  \tilde{G}^{cd}\right]  & 0 & 0\\
0 & 0 & \frac{J}{m} & 0\\
0 & 0 & 0 & \frac{J}{m}
\end{array}
\right],
\end{equation}
where
\[ 
\left[
  \tilde{G}^{cd}\right]  =\left[
\begin{array}
[c]{ccc}
\frac{1}{R^{2}\sin^{2}\left(  \frac{r}{R}\right)  } & \frac{-\cos\left(
\frac{r}{R}\right)  }{R^{2}\sin^{2}\left(  \frac{r}{R}\right)  } & \frac
{-\cos\left(  \frac{r}{R}\right)  }{R^{2}\sin^{2}\left(  \frac{r}{R}\right)}\\{}&{}&{}\\
\frac{-\cos\left(  \frac{r}{R}\right)  }{R^{2}\sin^{2}\left(  \frac{r}
{R}\right)  } & \frac{m}{Jx^{2}}+\frac{-\cos^{2}\left(  \frac{r}{R}\right)
}{R^{2}\sin^{2}\left(  \frac{r}{R}\right)  } & \frac{\cos^{2}\left(  \frac
{r}{R}\right)  }{R^{2}\sin^{2}\left(  \frac{r}{R}\right)  }\\{}&{}&{}\\
\frac{-\cos\left(  \frac{r}{R}\right)  }{R^{2}\sin^{2}\left(
\frac{r} {R}\right)  } & \frac{\cos^{2}\left(
\frac{r}{R}\right)  }{R^{2}\sin ^{2}\left(  \frac{r}{R}\right)  }
& \frac{m}{Jy^{2}}+\frac{-\cos^{2}\left( \frac{r}{R}\right)
}{R^{2}\sin^{2}\left(  \frac{r}{R}\right)  }
\end{array}
\right].
\]
In the pseudospherical case the analogous equations are valid.
The only difference is that trigonometric functions are replaced
by (properly signed) hyperbolic functions. So we have
$T=\frac{m}{2}G_{ij}\frac{dq^{i}}{dt}\frac{dq^{j}}{dt}$, where
$\left[\dots q^{i}\dots \right]
=\left[r,\varphi,\gamma,\delta,x,y\right]$\ and the matrices
$[G_{ij}]$\ and $\left[  G^{ab}\right]$\ are given in the same form as in spherical problem
(\ref{two 208}), (\ref{two 209}). We must only replace function $\sin\frac{r}{R}$ by
$\sinh\frac{r}{R}$\ and $\cos\frac{r}{R}$\ by $\cosh\frac{r}{R}$.
In the canonical formalism it is similarly.  The generalized coordinates are
ordered as $\left[r,\varphi,\gamma,\delta,x,y\right]$, $r$\ is
now "pseudospherical radius" and $r\in \left(  0...\infty\right)$.
Explicitly the kinetic term in the Hamiltonian on the sphere has the form
\begin{eqnarray}
\nonumber
T&=&\frac{{p_{r}}^{2}}{2m}+\frac{p_{\varphi }^{2}-2\cos \frac{r}{R}p_{\varphi }\left(
p_{\gamma }+p_{\delta }\right) }{2mR^{2}\sin ^{2}\frac{r}{R}}\\
&+&
\label{two 210}\frac{ \left(
\frac{mR^{2}}{J}\sin ^{2}\frac{r}{R}+\cos ^{2}\frac{r}{R}\right) \left( p_{\gamma
}+p_{\delta }\right) ^{2}}{2mR^{2}\sin ^{2}\frac{r}{R}}\\
\nonumber
 &+& \frac{p_{x}^{2}}{2J}+\frac{p_{y}^{2}}{2J}+\frac{p_{\gamma }^{2}}{2Jx^{2}}+
\frac{p_{\delta }^{2}}{2Jy^{2}},
\end{eqnarray}
and on the pseudosphere
\begin{eqnarray}
\nonumber
T&=&\frac{{p_{r}}^{2}}{2m}+\frac{p_{\varphi }^{2}-2\cosh \frac{r}{R}p_{\varphi }\left(
p_{\gamma }+p_{\delta }\right) }{2mR^{2}\sinh ^{2}\frac{r}{R}}\\
\label{two 211}
&+&\frac{ \left( \cosh
^{2}\frac{r}{R}\pm \frac{mR^{2}}{J}\sinh ^{2}\frac{r}{R}\right) \left( p_{\gamma
}+p_{\delta }\right) ^{2}}{2mR^{2}\sinh ^{2}\frac{r}{R}} 
\\
\nonumber
&+&\frac{p_{x}^{2}}{2J}+\frac{p_{y}^{2}}{2J}+\frac{p_{\gamma }^{2}}{2Jx^{2}}+
\frac{p_{\delta }^{2}}{2Jy^{2}}.
\end{eqnarray}
The Hamilton-Jacobi equation: $\frac{\partial S}{\partial t}+H\left( q,\frac{\partial
S}{\partial q}\right) =0$ will be again reduced by the substitution:
$S=-Et+S_{0}\left( q\right).$ The time-independent Hamilton-Jacobi equation has the form
$H\left( q,\frac{\partial S_{0}}{\partial q}\right) =E$, and we seek for solutions of the form $S_{0}=S_{r}\left( r\right)
+S_{\varphi }(\varphi )+S_{\gamma }(\gamma )+S_{\delta }(\delta )+S_{x}(x)+S_{y}(y)
=S_{r}(r)+l\varphi +C_{\gamma }\gamma +C_{\delta }\delta +S_{x}(x)+S_{y}(y)$. We can
separate these equations if $\varphi$, $\gamma $ and $\delta $ are cyclic  variables.
It is more convenient if we put $p_{\gamma }=\frac{1}{2}\left( p_{\alpha }+p_{\beta
}\right) $ and $p_{\delta }=\frac{1}{2}\left( p_{\alpha }-p_{\beta }\right).$ Then:
$S_{0}=S_{r}\left( r\right) +S_{\varphi }(\varphi )+S_{\gamma }(\gamma )+S_{\delta
}(\delta )+S_{x}(x)+S_{y}(y)=S_{r}(r)+l\varphi +C_{\alpha }\alpha +C_{\beta }\beta
+S_{x}(x)+S_{y}(y).$
In the spherical case the Hamilton-Jacobi equation has the form: 
\begin{eqnarray}
\nonumber
E&=&\frac{1}{2m}\left( \frac{d S_{r}\left( r\right) }{d r} \right)
^{2}+\frac{\left( l-C_{\alpha }\cos \frac{r}{R}\right) ^{2}}{ 2mR^{2}\sin
^{2}\frac{r}{R}}+V(r) \\
&+& \frac{1}{2J}\left( \frac{d S_{x}}{d x} \right)
+\frac{\left( C_{\alpha }+C_{\beta }\right) ^{2}}{8Jx^{2}}+V_{x}(x) \\
\nonumber
&+&\frac{1}{2J}\left( \frac{d S_{y}}{d y}\right) +\frac{\left( C_{\alpha
}-C_{\beta }\right) ^{2}}{8Jy^{2}}+V_{y}(y).
\end{eqnarray}
In the pseudospherical case: 
\begin{eqnarray}
\nonumber
 E&=&\frac{1}{2m}\left( \frac{d S_{r}\left(
r\right) }{d r}\right) ^{2}+\frac{\left( l-C_{\alpha }\cosh \frac{r}{R}\right)
^{2}}{ 2mR^{2}\sinh ^{2}\frac{r}{R}}+V(r) \\
&+&\frac{1}{2J}\left( \frac{d
S_{x}}{d x} \right)+ \frac{\left( C_{\alpha }+C_{\beta }\right)
^{2}}{8Jx^{2}}+V_{x}(x)  \\
\nonumber
&+&\frac{1}{2J}\left( \frac{d S_{y}}{d y}\right)
+\frac{\left( C_{\alpha }-C_{\beta }\right) ^{2}}{8Jy^{2}}+V_{y}(y).
\end{eqnarray}
Now we can calculate the action variables. For the all considered cases $ \alpha
,\beta $ and $\delta $ are cyclic variables and the corresponding actions have the
same form:
\begin{eqnarray}
\nonumber
J_{\varphi }&=&\oint \frac{d S_{\varphi }(\varphi )
}{d \varphi }d\varphi =\int_{0}^{2\pi }ld\varphi =2\pi l\Rightarrow l=\frac{
J_{\varphi }}{2\pi },
\\ \label{two 212}
J_{\alpha }&=&\oint \frac{d S_{\alpha }(\alpha )}{
d \alpha }d\alpha =C_{\alpha }\int_{0}^{2\pi }d\alpha =2\pi C_{\alpha }\Rightarrow
C_{\alpha }=\frac{J_{\alpha }}{2\pi },
\\ \nonumber
J_{\beta }&=&\oint \frac{d S_{\beta }(\beta )}{d
\beta } d\varphi =C_{\beta }\int_{0}^{2\pi }d\beta =2\pi C_{\beta }\Rightarrow
C_{\beta}=\frac{J_{\beta }}{2\pi }.
\end{eqnarray}
Now we can simplify the Hamilton-Jacobi equation using $J_{\varphi }$, $ J_{\alpha
}$, $J_{\beta }$. After this  we can calculate $J_{r},J_{x}$, $J_{y}$. The constants
of separation $C_{x}$ and $C_{y}$   have the same form for the sphere and pseudosphere cases:
\begin{eqnarray}
&&C_{x}:=\frac{1}{2J}\left( \frac{d S_{x}}{d x}\right) +\frac{\left(
J_{\alpha }+J_{\beta }\right) ^{2}}{32\pi ^{2}Jx^{2}}+V_{x}(x), \\
&&C_{y}:=\frac{1}{2J}\left( \frac{d S_{y}}{d y}\right) +\frac{\left(
J_{\alpha }-J_{\beta }\right) ^{2}}{32\pi ^{2}Jy^{2}}+V_{y}(y).
\end{eqnarray}
1. On the sphere we obtain:
\begin{eqnarray}
\nonumber
J_{r}&=&\oint \sqrt{2m\left( E-C_{x}-C_{y}-V_{r}(r)\right) - \frac{\left(
J_{\varphi }-J_{\alpha }\cos \frac{r}{R}\right) ^{2}}{4\pi ^{2}R^{2}\sin
^{2}\frac{r}{R}}}dr,
\\
 \label{two 213}
J_{x}&=&\oint \sqrt{2J\left( C_{x}-V_{x}(x)\right) -\frac{ \left(
J_{\alpha }+J_{\beta }\right) ^{2}}{16\pi ^{2}x^{2}}}dx,
\\ \nonumber
J_{y}&=&\oint \sqrt{2J\left( C_{y}-V_{y}(y)\right) -\frac{ \left(
J_{\alpha }-J_{\beta }\right) ^{2}}{16\pi ^{2}y^{2}}}dy.
\end{eqnarray}
2. On the pseudosphere the formulas read:
\begin{eqnarray}
\nonumber
J_{r}&=&\oint \sqrt{2m\left( E-C_{x}-C_{y}-V_{r}(r)\right) - \frac{\left(
J_{\varphi }-J_{\alpha }\cosh \frac{r}{R}\right) ^{2}}{4\pi ^{2}R^{2}\sinh
^{2}\frac{r}{R}}}dr,
\\ \label{two 214}
J_{x}&=&\oint \sqrt{2J\left( C_{x}-V_{x}(x)\right) -\frac{ \left(
J_{\alpha }+J_{\beta }\right) ^{2}}{16\pi ^{2}x^{2}}}dx,
\\ \nonumber
J_{y}&=&\oint \sqrt{2J\left( C_{y}-V_{y}(y)\right) -\frac{ \left(
J_{\alpha }-J_{\beta }\right) ^{2}}{16\pi ^{2}y^{2}}}dy.
\end{eqnarray}
Substituting (\ref{two 212}) into (\ref{two 212}), (\ref{two 214}) and expressions for $C_{x}$ and
$C_{y}$ one obtains the explicit dependence of $E$ on the action variables.
  The Hamilton-Jacobi equation is separable for some realistic (in the elasticity theory sense)
potentials of the form $\mathfrak{U} (q)=V_{r}(r)+V_{x}(x)+V_{y}(y)$ both on the sphere and pseudosphere.
\bigskip
\subsection{ Models of potentials}
\bigskip
\subsubsection{ Models of potentials in the "deformations" plane }
\bigskip
There exists an interesting class of  "universally" separable potentials. The corresponding Hamilton-Jacobi equations are separable in all coordinate systems used below, i.e., Cartesian coordinates $(x,y)$ and polar coordinates $(\varsigma$, $\varepsilon)$ in the $(x,y)$-plane.
Among those potentials there are ones effectively integrable, realistic, and applicable in elastic problems concerning internal degrees of freedom. Some of them  are also separable in elliptic coordinates in the $(x,y)$-plane, however, they are rather non-useful in elastic problems, so we do not make use of elliptic coordinates and corresponding action variables. The mentioned potentials have the general form:
\begin{equation}
V(x,y)=\frac{A}{x^{2}}+\frac{B}{y^{2}}+C\left( x^{2}+y^{2}\right) ,
\end{equation}
where $A$, $B$, $C$ are constants. An interesting and applicable subclass is given by: 
\begin{equation}
V\left( x,y\right) =\frac{F}{y^{2}}+\frac{F}{4}\left( x^{2}+y^{2}\right),
\end{equation}
$F$ is constant. The action variables are then given by 
\begin{equation*}
J_{x}=-\sqrt{\frac{(J_{\alpha }+J_{\beta })^{2}}{16}}-\pi C_{x}\sqrt{\frac{2I%
}{F}},\quad J_{y}=-\pi C_{y}\sqrt{\frac{2I}{F}}-\pi \sqrt{2IF+\frac{%
(J_{\alpha }-J_{\beta })^{2}}{16\pi ^{2}}}.
\end{equation*}
Another class of separable potentials very well suited to nonlinear elastic problems is given by expressions:
\begin{eqnarray*}
V(\varsigma ,\varepsilon )=V_{\varsigma }(\varsigma )+\frac{V_{\varepsilon }(\varepsilon )}{\varsigma ^{2}\sin
^{2}(\varepsilon )},
\end{eqnarray*}
where $(\varsigma , \varepsilon)$ are polar coordinates in the $(x,y)$-plane of deformation invariants:
\begin{equation}
x=\varsigma \sin
\varepsilon ,\qquad y=\varsigma \cos \varepsilon ,  
\end{equation}
$V_{\varsigma }(\varsigma )$, $V_{\varepsilon }(\varepsilon )$ are functions of the one indicated variable, respectively $\varsigma$, $\varepsilon$. 
Appropriately choosing 
these "shape functions" we can just obtain the 
mentioned compatibility with the standard 
requirements of nonlinear elastic dynamics. 
 In spherical case (we mean here spherical 
geometry of two-dimensional "physical" space) 
when generalized coordinates are ordered as 
$[r,\varphi , \alpha , \beta , \varepsilon , \varepsilon]$ , the metric $\left[ G^{ij}\right],$ has the form
\begin{equation}
\left[ G^{ij}\right]=\left[
\begin{array}{cccc}
1&0&0&0\\
0&\left[ \tilde{G}^{kl}\right]&0&0\\
0&0&\frac{m}{\varsigma ^{2}J}&0\\
0&0&0&\frac{m}{J}
\end{array} \right]
\end{equation}
and
\[
\left[
\tilde{G}^{kl}\right]=\left[
\begin{array}{ccc}
{\frac{1}{R^{2}\sin ^{2}\left( \frac{r}{R}\right) }} & {\frac{-\cos \left(
\frac{r}{R}\right) }{R^{2}\sin ^{2}\left( \frac{r}{R}\right) }} &  {\frac{ -\cos
\left( \frac{r}{R}\right) }{R^{2}\sin ^{2}\left( \frac{r}{R}\right) }}
\\{}&{}&{}\\
{ {\frac{-\cos \left( \frac{r}{R}\right) }{R^{2}\sin ^{2}\left( \frac{r}{R} \right)
}}} & { { {\wp}+\frac{\cot ^{2}\left( \frac{r }{R}\right)}{R^{2}} }} &
 { {\frac{1}{R^{2}}\cot ^{2}\left( \frac{r
}{R}\right) }} \\{}&{}&{}\\
 {\frac{-\cos \left( \frac{r}{R}\right) }{R^{2}\sin ^{2}\left( \frac{r}{R} \right)
}} & { {\frac{1}{R^{2}}\cot ^{2}\left( \frac{r}{R}\right) }}& {{ \wp+\frac{\cot
^{2}\left( \frac{r }{R}\right)}{R^{2}} }}
\end{array}
\right]
\],
where 
\begin{equation}
\wp:={\frac {m}{J{(\varsigma \cos \varepsilon )}^{2}}}.
\end{equation}
The separation constants $C_{x}$, $C_{y}$ will be combined into a single one,
\begin{equation}
C=-C_{x}-C_{y}
\end{equation}
 and
\begin{eqnarray}\nonumber
C&=&\frac{1}{2J}\left( \frac{\partial S_{\varsigma }}{\partial \varsigma }%
\right) ^{2}+\frac{1}{2J\varsigma ^{2}}\left( \frac{\partial S_{\varepsilon }%
}{\partial \varepsilon }\right) ^{2}
\\ 
\label{two 215}
&+&\frac{\left( J_{\alpha }+J_{\beta
}\right) ^{2}}{8\pi ^{2}J\varsigma ^{2}(\sin 2\varepsilon )}+V_{\varsigma
\varepsilon }(\varsigma ,\varepsilon ).
\end{eqnarray}
The new constant of separation $A$ appears
\begin{eqnarray}
A&:=&\frac{1}{2J}\left( \frac{\partial S_{\varepsilon }}{\partial \varepsilon }%
\right) ^{2}\frac{\left( J_{\alpha }+J_{\beta }\right) ^{2}}{8\pi
 ^{2}J(\sin
2\varepsilon )}+V_{\varepsilon }(\varepsilon ),\\
J_{\varepsilon }&=&\oint \sqrt{2J\left( A-V_{\varepsilon }(\varepsilon
)\right) -\frac{\left( J_{\alpha }^{2}+2\cos (2\varepsilon )J_{\alpha
}J_{\beta }+J_{\beta }^{2}\right) }{4\pi ^{2}R^{2}\sin ^{2}(2\varepsilon )}}%
d\varepsilon ,\\
J_{\varsigma }&=&\oint \sqrt{2I\left( C-V_{\varsigma }(\varsigma )\right) -%
\frac{2JA}{\varsigma ^{2}}}d\varsigma .
\end{eqnarray}
As $V_{\varepsilon }(\varepsilon )$ we propose:
\[V_{\varepsilon }(\varepsilon
)=\widehat{\gamma }\mathrm{\cot }^{2}(2\varepsilon ),\]
where $\widehat{\gamma}$ is some constant, then:
\begin{equation}
J_{\varepsilon }=\frac{1}{4}\left( 4\sqrt{2J(A+\widehat{\gamma })}\pi - \sqrt{8J\widehat{\gamma }\pi ^{2}+(J_{\alpha }-J_{\beta })^{2}}-
\sqrt{8J\widehat{\gamma }\pi ^{2}+(J_{\alpha }+J_{\beta })^{2}}\right). \nonumber 
\end{equation}
As $V_{\varsigma }(\varsigma )$ let us take 
\[V_{\varsigma }(\varsigma )=
\frac{\widetilde{\gamma }}{\varsigma },
\] 
where $\widetilde{ \gamma}$ is some constant,then:
\begin{equation}
J_{\varsigma }=\sqrt{2}\left( -2\sqrt{JA}+\frac{\sqrt{J}\widetilde{\gamma }}{%
\sqrt{C_{x}+C_{y}}}\right) \pi . \nonumber
\end{equation}
\bigskip
\subsubsection{ Models for the potential $V_{r}(r)$}
\bigskip
The next important problem is to suggest 
some physically interesting and computationally 
effective models for the "radial" potentials $V_{r}(r)$ 
in the physical space. One can expect that 
computationally effective will be potentials somehow 
 suited  to geometry of the "physical" space $M$. 
It is well-known that the determinant ${\rm det} [g^{ij}] $ 
is a scalar density of weight $-2$.
This is an important geometric  
object, used, e.g., for expressing the two-dimensional 
"volume" (surface area) element. Namely, this 
element is analytically given by
\begin{equation}
d\mu (r,\varphi )\frac{1}{\sqrt{{\rm det}[g^{ij}]}}\, dr d\varphi = {\sqrt{{\rm det}[g_{ij}]}}\, dr d\varphi  .
\end{equation}
One can expect that simple expresions built 
of ${\rm det} [g^{ij}]$ will be good candidates for the 
radial potentials, 
both mechanically interesting and computationally 
effective. We assume the general 
form:
\begin{equation} \label{two 216}
 V_{r}(r)=f(r){\rm det}[g^{ij}].
\end{equation}
$f$ being something "simple". Obviously, it is 
constant that is the simplest; we denote it by
\[
f=R^{2}\gamma ,
\]
 $\gamma $ being also some constant, and then 
in spherical case where ${\rm det}[g_{ij}]=\frac{1}{R^{2}\sin^{2}(\frac{r}{R})}$ we have potential $V_{r}(r)$ in the form:
\begin{equation*}
V_{r}(r)=\frac{\gamma }{\sin ^{2}(\frac{r}{R})}.
\end{equation*}
The "radial" action variable $J_{r}$ (\ref{two 212}) is then given by:
\begin{eqnarray}\nonumber
J_{r} &=&\sqrt{2m4\pi ^{2}R^{2}(E+C)+J_{\alpha }^{2}}\\
&-& \frac{1}{2}%
\sqrt{2m4\pi ^{2}R^{2}\gamma +(J_{\varphi }+J_{\alpha })^{2}}\label{two 217} 
\\ \nonumber
&-&\frac{1}{2}\sqrt{2m4\pi ^{2}R^{2}\gamma +(J_{\varphi }-J_{\alpha })^{2}}.
\end{eqnarray}
With $-C_{x}-C_{y}=C$ substituted from (\ref{two 215}). 
Obviously, in the compact spherical manifold 
all translational  motion are bounded on the 
"physical" space, therefore, the situation where $\gamma=0$, 
i.e., $V_{r}(r)=0$ are admissible for the 
existence of well-defined action variables $J_{r}$. 
then we have 
\begin{eqnarray}\nonumber
J_{r}&=&\sqrt{2m4\pi ^{2}R^{2}( E-C_{x}-C_{y})+J_{\alpha }^{2}}\\
\label{two 218}
&-&\frac{1}{2}%
|J_{\varphi }-J_{\alpha }| -\frac{1}{2}|J_{\varphi }-J_{\alpha }|
=\wr J_{\varphi } \geq |J_{\alpha }|\wr \\ \nonumber &=&\sqrt{2m4\pi
^{2}R^{2}(E-C_{x}-C_{y})+J_{\alpha }^{2}}-J_{\varphi }.
\end{eqnarray}
\bigskip
In pseudospherical case
\[{\rm det}[g_{ij}]=\frac{1}{R^{2}\sinh^{2}(\frac{r}{R})}\] 
and again we put
\[f(r)=\gamma,\]
$\gamma$ being constant, and then 
\begin{equation*}
V(r)=\frac{\gamma }{R^{2}\sinh ^{2}(\frac{r}{R})},
\end{equation*}
This also leads to some explicitly 
integrable expression for $J_{r}$ (\ref{two 214}). However 
negative if motion in the $r$-variable 
is to be bounded and the corresponding 
action variable $J_{r}$ to be finite.  
\begin{eqnarray*}
J_{r}&=&-\sqrt{2m4\pi^{2}R^{2}(C_{x}\!+\!C_{y}-\!E )\!+\!J_{\alpha}^{2}}
\\
&+&\frac{1}{2}\sqrt{2m4\pi^{2}\gamma\!+\!(J_{\varphi}\!+\!J_{\alpha})^{2}}
\\&-& \frac{1}{2}{\sqrt{2m4\pi^{2}\gamma +(J_{\varphi}-J_{\alpha})^{2}}}.
\end{eqnarray*}
\newline
Solving the formulas (\ref{two 217}), (\ref{two 218})with 
respect to $E$, we express the energy  
as a function of all action variables. For that 
$C_{x}$, $C_{y}$ must be also expressed by action 
variables. One achieves this by solving 
(\ref{two 212}) with respect to  $C_{x}$, $C_{y}$ 
and substituting the resulting expressions 
to (\ref{two 218}), (\ref{two 218}).
\bigskip
\section{Some additional remarks}
\bigskip 
We have mentioned above that there exist some models, 
interesting at least from purely geometric point of view of 
analytical mechanics in itself. This has to do with our 
models of affinely invariant dynamics as developed in  
\cite{JJSco2004}, \cite{IJJSco2004}, \cite{IIJJSco2005}.
Namely, we can assume in $M$ only some affine connection 
structure $(M,\Gamma)$ without any fixed metric tensor $g$ on $M$. 
The role of instantaneous metric tensor at the configuration 
$e\in F_{x}M \subset FM$ will be played by the Cauchy deformation 
tensor; analytically
\begin{equation}\
C[e]_{ij}=\eta_{AB}e^{A}{}_{i}e^{B}{}_{j},
\end{equation}
 where $\eta$ denotes the micromaterial metric. One can always 
choose coordinates in such a way that $\eta_{AB}=\delta_{AB}$ (if 
$\eta$ is positively definite, what is physically assumed). And further 
on we postulate kinetic energy in the "usual" form:
\begin{eqnarray}
T&=&\frac{m}{2}C_{ij}[e]\frac{dx^{i}}{dt}\frac{dx^{j}}{dt}+\frac{1}{2} 
C_{ij}[e] \left( \frac{D}{Dt} e^{i}{}_{A}\right)  \left( \frac{D}{Dt}e^{j}{}_{A}\right)J^{AB}\\
&=& \frac{m}{2}\eta_{AB}\widehat{v}^{A}\widehat{v}^{B}+
\frac{1}{2}\eta_{AB}\widehat{\Omega}^{A}{}_{C}\widehat{\Omega}^{B}{}_{D}J^{CD},
\end{eqnarray}
where $\widehat{v}^{A}$ are co-moving components of translational 
velocity. After some calculations based on Poisson brackets 
we obtain equations of motion in the following balance form:
\begin{eqnarray}
m\frac{Dv^{a}}{Dt}&=&m\left(\Omega^{a}{}_{b}+\Omega_{b}{}^{a}\right) v^{b}
+2mv^{b}v^{c}S_{cb}{}^{a}+\Sigma^{c}{}_{d}R^{d}{}_{cab}v^{b}+F^{a},\\
m\frac{D\Sigma^{ab}}{Dt}&=&\Sigma ^{ac} \left(\Omega_{c}{}^{b}+\Omega^{b}{}_{c}\right) - m v^{a}
v^{b}+N^{ab},
\end{eqnarray}
where indices are raised and lowered with the help of $C$ 
as a "metric tensor". In the purely covariant and mixed tensor 
form, these equations acquire the following suggestive shape:
\begin{eqnarray}
\frac{Dp_{a}}{Dt}=m\frac{Dv_{a}}{Dt}&=&2mv^{b}v^{c}S_{cba}+\Sigma^{c}{}_{d}R^{d}{}_{cab}v^{b}+F^{a},\\
m\frac{D\Sigma^{a}{}_{b}}{Dt}&=&N^{a}{}_{b}- m v^{a}
v_{b},
\end{eqnarray}
with the connection concerning the shift of indices. 
These formulas are very similar formally to (\ref{eq 152}),
however we must remember about important difference:
no metric field $g$ is now defined all over $M$. The Cauchy "metric" 
$C[e]$ is defined only at the point $x=\pi (e)$ where the object 
is instantaneously present.
\newline
Although no metric field $g$ is fixed in $M$, nevertheless 
we can introduce the concept of rigid motion by putting 
the following conditions on motion of our object:
\begin{equation}
\frac{DC_{ij}}{Dt}=0,
\end{equation}
and this is simply equivalent to:
\begin{equation}
\Omega^{i}{}_{j}=-\Omega_{j}{}^{i}=-C_{jk}{C^{-1}}^{il}\Omega^{k}{}_{l},
\end{equation}
i.e., the $C$--skew--symmetry of $\Omega$. It is interesting that 
unlike the usual $g$-skew-symmetry of $\Omega$, i.e., the $g$-metrical
rigid motion, these constrains are in general non-holonomic.
\newline
To finish with let us remind our models of affinely--invariant 
dynamics (in the left- and rigid--sense) of extended affine bodies 
in flat spaces. There are natural in counterparts manifolds.
Here we merely quote the corresponding kinetic energies, analogous 
to ones introduced in \cite{JJSco2004}, \cite{IJJSco2004}, \cite{IIJJSco2005}.
\newline
For the spatially affine and micromaterially isotropic models
we have:
\begin{eqnarray}\nonumber
T&=&\frac{m}{2} \eta_{AB} \widehat{v}^{A}\widehat{v}^{B}+
\frac{I}{2}\eta_{KL}\eta^{MN}
\widehat{\Omega}^{K}{}_{M}\widehat{\Omega}^{L}{}_{N}+
\frac{A}{2}{\rm Tr} \left( \widehat{\Omega}^{2} \right)+
\frac{B}{2}{\rm Tr} \left( \widehat{\Omega} \right)^{2}\\ \nonumber
&=&\frac{m}{2} C_{ij} {v}^{i} {v}^{j}+
\frac{I}{2}C_{kl}C^{mn}
 {\Omega}^{k}{}_{m} {\Omega}^{l}{}_{n}+
\frac{A}{2}\Omega^{i}{}_{j} \Omega^{j}{}_{i}+
\frac{B}{2}\Omega^{i}{}_{i} \Omega^{j}{}_{j}.
\end{eqnarray}
Similarly, for the spatially metrical and micromaterially affine 
models (discretization of the Arnold description of fluids) we have:
\begin{eqnarray}\nonumber
T&=&\frac{m}{2} G_{AB} \widehat{v}^{A}\widehat{v}^{B}+
\frac{I}{2}G_{KL}G^{MN}
\widehat{\Omega}^{K}{}_{M}\widehat{\Omega}^{L}{}_{N}+
\frac{A}{2}{\rm Tr} \left( \widehat{\Omega}^{2} \right)+
\frac{B}{2}{\rm Tr} \left( \widehat{\Omega} \right)^{2}\\\nonumber
&=&\frac{m}{2} g_{ij} {v}^{i} {v}^{j}+
\frac{I}{2}g_{kl}g^{mn}
 {\Omega}^{k}{}_{m} {\Omega}^{l}{}_{n}+
\frac{A}{2}\Omega^{i}{}_{j} \Omega^{j}{}_{i}+
\frac{B}{2}\Omega^{i}{}_{i} \Omega^{j}{}_{j}.
\end{eqnarray}
Obviously, in spite of this redundant way of writing, we remember 
that ${\rm Tr}\left( \Omega^{2}\right)={\rm Tr}\left( \widehat{\Omega}^{2}\right)$, ${\rm Tr}\left( \Omega \right)={\rm Tr}\left( \widehat{\Omega}\right)$.
Models of this kind were analyzed in flat space motion
\cite{JJSco2004}, \cite{IJJSco2004}, \cite{IIJJSco2005}. In curved manifolds the problem is much more 
complicated and will be analysed only in some special cases. The more 
detailed study is postponed to the later papers.
\newline
{\bf{Acknowledgment}}
The research presented above was supported by the Ministry of Science and Higher Education grant No 501 018 32/1992.


\bigskip
\begin{thebibliography}{99}
\bibitem{Abr1978} 
R. Abraham and J. E. Marsden, {\em Foundations of Mechanics} (second
ed.), The Benjamin-Cummings Publishing Company, London-Amsterdam-
Sydney-Tokyo, 1978.
\bibitem{Arn1978} 
V. I. Arnold, {\em Mathematical Methods of Classical Mechanics}, Springer Grad-
uate Texts in Mathematics, vol. {\bf 60}, Springer-Verlag, New York, 1978.
\bibitem{Bloo1979} 
F. Bloom, {\em Modern Differential Geometric Techniques in The Theory of Continuous Distributions of Dislocations}, Lecture Notes in Math., vol. {\bf 733}, Springer-Verlag, Berlin, (1979).
\bibitem{Bor2004} 
A. V. Borisov, I. S. Mamaev, {\em Classical dynamics in non-Eucledian spaces}.
/Sb. statei, Moscow - Izhevsk: Institute of Computer Science, (2004), 348
p. (in Russian).
\bibitem{Bress1978} 
A. Bressan, {\em Relativistic theories of materials}, Springer-Verlag, New York,
1978.
\bibitem{Bur2008} 
A. A. Burov, D. P. Chevallier, {\em  Dynamics of Affinely Deformable Bodies From the Stand Point of Theoretical Mechanics and Differential Geometry}, Accepted to Reports on Mathematical Physics, (2008)
\bibitem{2Bur2008} 
A. A. Burov, {\em The motion of a body with a plane of symmetry over a three-
dimensional sphere under the action of a spherical analogue of Newtonian
gravitationstar} Journal of Applied Mathematics and Mechanics Vol. {\bf 72},
2008, p. 15--21.
\bibitem{Bur1996} 
A. A. Burov and D. P. Chevallier, {\em On the Variational Principle of Poincare,
the Poincare-Chetayev Equations and the Dynamics of Affinely Deformable
Bodies}, Cahier de C.E.R.M.I.C.S. {\bf 14}, Mai 1996.
\bibitem{Bur1995} 
A. A. Burov and S. Ya. Stepanov, {\em On Geometry of Masses in Dynamics of
Deformable Bodies}, in: Problems of Investigation on Stability and Stabi-
lization of Motion, Computing Centre of the Russian Academy of Sciences,
Moscow, 1995 (in Russian).
\bibitem{Cap2003} 
G. Capriz and P. M. Mariano, {\em Symmetries and Hamiltonian Formalism for
Complex Materials}, Journal of Elasticity {\bf 72} (2003), 57--70.
\bibitem{Cap1989} 
G. Capriz, {\em Continua with Microstructure}, New York : Springer-Verlag,
 1989.
\bibitem{Cas1995} 
J. Casey, {\em On the Advantages of a Geometrical Viewpoint in the Derivation of Lagrange's Equations for a Rigid Continuum}, Theoretical, Experimental and Numerical Contributions to the Mechanics of Fluids and Solids, Special Issue of Journal of Applied Mechanics and Physics, {\bf 46}, (1995), S805--S847.
\bibitem{Che2004} 
D. P. Chevallier, {\em On the Foundations of Ordinary and Generalized Rigid
Body Dynamics and the Principle of Objectivity}, Arch. Mech. {\bf 56} (2004),
no. 4, 313--353.
\bibitem{Coss1909} 
E. Cosserat, F. {\em Cosserat Th\'{e}orie des crops d\'{e}formables}, A. Hermann et Fils, Paris, 1909.
\bibitem{Coss1908} 
E. Cosserat, F. Cosserat,{\em Sur la Th\'{e}orie des corps minces}, Compt. Rend., 146, 1908, pp. 169--172.
\bibitem{2Coss1909} 
E. Cosserat, F. Cosserat, {\em Sur la Th\'{e}orie de L'elasticite}, Ann. Toulouse, 1909.
\bibitem{Erin1968} 
A. C. Eringen, {\em  Mechanics of Micromorphic Continua}, in: Proceedings of
the IUTAM Symposium on Mechanics of Generalized Continua, Freuden-
stadt and Stuttgart, 1967, editor: E. Kr\"{o}ner, vol. 18, Springer, Berlin-
Heidelberg-New York, 1968, 18--33.
\bibitem{Erin64} 
A. C. Eringen, {\em Nonlinear Theory of Micro Elastic Solids}, Part I and II Int. J. Eng. Sci., (1964).
\bibitem{Erin1962} 
A. C. Eringen, {\em Nonlinear Theory of Continuous Media}, McGraw-Hill Book
Company, New York, 1962.
\bibitem{Gol2004} B. Go\l ubowska, {\em Action-Angle Analysis of Some Geometric Models of Internal Degrees of Freedom}, J. of Nonlinear Math. Phys. {\bf 11} (2004), Supplement, 138-144.
\bibitem{Gol2006} 
B. Go\l ubowska, {\em An Affinely-Rigid Body in Manifolds and Spaces with the
Constant Curvature}, PhD thesis, Warsaw, 2006.
\bibitem{Gol1960} 
I. I. Goldenblatt, {\em Nonlinear Problems in Elasticity}, (in Russia), Moskow, Nauka 1960.
\bibitem{Kadi1983} 
A. Kadi\u{c}, D. G. B. Edelen, {\em A Gauge Theory of Dislocations and Disclinations}, Springer-Verlag, Berlin--Heidenberg--New York, 1983.
\bibitem{Kan2004} 
E. Kanso, P. Papadopoulos, {\em Pseudo-Rigid Ball Impact on an Oscillating Rigid Foundation}, International Journal of Non-Linear Mechanics {\bf 39} (2004) 1129--1145.
\bibitem{Kob1963} 
S. Kobayashi and K. Nomizu,{\em Foundations of Differential Geometry}, In-
terscience Publishers, New York, 1963.
\bibitem{Kun1972}
H. P. K\"{u}nzle, Commun. Math. Phys. {\bf 27}, 23, 1972.
\bibitem{Lew1990} 
D. Lewis, J. C. Simo, {\em Nonlinear Stability of Rotating Pseudo Rigid Body}, Proc. Roy. Soc. A 1873, {\bf 427}, 1990, 281 p.
\bibitem{Mar2000} 
P. M. Mariano, {\em Configuration Forces in Continua with Microstructure}, Z.
angew. Math. Phys. {\bf 51} (2000), 752-791. 
\bibitem{Mars1994} 
J. E. Marsden and T. Ratiu, {\em Introduction to Mechanics and Symmetry},
Springer, New York, 1994.
\bibitem{Mars1999} 
J. E. Marsden and T. Ratiu, {\em Introduction to Mechanics and Symmetry}. A
Basic Exposition of Classical Mechanical Systems (second ed.), Springer,
New York, 1999.
\bibitem{Mars1983} 
J. E. Marsden, T. J. R. Hughes, {\em Mathematical Foundations of Elasticity},
Prentice-Hall Engelewood Cliffs, N.J. (1983).
\bibitem{51Math} 
M. Mathison, Acta Phys. Polon. {\bf 6}, 163, (1967).
\bibitem{Mig2003} 
A. San Miguel, {\em Stability of motion by similarity transformations with
a fixed point} Monograf\'{i}as del Semin. Matem. Garc\'{i}a de Galdeano 27:
515521, 2003.
\bibitem{5Mig2003} 
A. San Miguel, {\em Deformable Asymmetric Tops under Similarity Transfor-
mations} Journal of Nonlinear Science. {\bf 13}, 5, 2003.
\bibitem{Now1070}
 W. Nowacki, {\em Teoria Niesymetrycznej Spr\c{e}\.{z}ysto\'{s}ci}, PAN, (1970), 246.
\bibitem{Now1968} 
W. Nowacki, {\em Couple Stress in The Theory of Termoelasticity}, Bull. Aced.
Polone. Sci., Ser. Sci. Techn. {bf 14}, 8, (1968).
\bibitem{Papa1951} 
A. Papapetrou, Proc. Roy. Soc. A. {\bf 109}, 248, 1951.
\bibitem{Papad2001} 
P. Papadopoulos, {\em On a Class of Higher-order Pseudo-rigid Bodies}, Math.
Mech. Sol., {\bf 6}, pp. 631--640, (2001).
\bibitem{Rei1998} 
O. M. O'Reilly and P. C. Varadi, {\em A Unified Treatment of Constraints in the
Theory of a Cosserat Point}, Journal of Applied Mathematics and Physics
(ZAMP), Vol. {\bf 49}, No. 2, pp. 205--223 (1998).
\bibitem{Rei1996} 
O. M. O'Reilly, {\em A Properly Invariant Theory of Infinitesimal Deformations
of an Elastic Cosserat Point}, Journal of Applied Mathematics and Physics
(ZAMP), Vol. {\bf 47}, No. 2, pp. 179--193 (1996).
\bibitem{Roz2005} 
E. E. Ro\.{z}ko, {\em Dynamics of Affinely-Rigid Bodies with Degenerate Dimen-
sion}, Rep. on Math. Phys. {\bf 56}, (2005), no. 3, 311--332.
\bibitem{Ros1998} 
G. Rosensteel, J. Troupe, {\em Nonlinear Collective Nuclear Motion}, arXiv:nucl-th/9801040v1, 20 Jan 1998.
\bibitem{Rub1985} 
M. B. Rubin, {\em On The Numerical Solution of One Dimensional Continuum
Problems Using The Theory of Cosserat Point} J. Appl. Mech. {\bf 52}, (1985), pp. 373--378.
\bibitem{2Rub1985} 
M. B. Rubin, {\em On The Theory of a Cosserat Point and Its Application to The
Numerical Solution of Continuum Problems}, J. Appl. Mech. {\bf 52} (1985), pp.
368--372.
\bibitem{Sch1965} 
A. J. Schild, J. A. Schlosser, J. Math. Phys. {\bf 6}, 1299 (1965).
\bibitem{Shch2000} 
A. V. Shchepetilov, {\em Two-Body Problem on Spaces of Constant Curvature. I.
Dependence of The Hamiltonian on The Symmetry Group and The Reduction
of The Classical System}. Theor. Math. Phys. {\bf 124}, (2000), pp. 1068--1981.
\bibitem{Simo1991} 
J. C. Simo, D. K. Lewis and J. E. Marsden, {\em Stability of Relative
Equilibria I: The Reduced Energy Momentum Method}, Arch. Rat. Mech.
Anal., Vol. {\bf 115}, 1991, pp. 15--59.
\bibitem{JJSco2004} 
J. J. S\l awianowski, V. Kovalchuk, A. S\l awianowska, B. Go\l ubowska,
A. Martens, E. E. Ro\.{z}ko, and Z. J. Zawistowski, {\em Invariant Geodetic Systems on Lie Groups and Affine Models of Internal and Collective Degrees
of Freedom}, Prace IPPT | IFTR Reports {\bf 7}, 2004.
\bibitem{IJJSco2004} 
J. J. S\l awianowski, V. Kovalchuk, A. S\l awianowska, B. Go\l ubowska,
A. Martens, E. E. Ro\.{z}ko, and Z. J. Zawistowski,{\em  Affine Symmetry in Mechanics of Collective and Internal Modes}. Part I. Classical Models, Rep. on
Math. Phys. {\bf 54} (2004), no. {\bf 3}, 373--427.
\bibitem{IIJJSco2005} 
J. J. S\l awianowski, V. Kovalchuk, A. S\l awianowska, B. Go\l ubowska,
A. Martens, E. E. Ro\.{z}ko, and Z. J. Zawistowski, {\em Affine Symmetry in Me-
chanics of Collective and Internal Modes}. Part II. Quantum Models, Rep.
on Math. Phys. {\bf 55} (2005), no. 1, 1--45.
\bibitem{Sousa1994} 
M. E. R Sousa Dias, {\em A Geometric Hamiltonian Approach for the Affine Rigid Body}, in P. Chossat eds., Bifurcation and Symmetry, 291--299, 1994,
Kluwer Academic Pub.
\bibitem{Sol2000} 
J. M. Solberg and P. Papadopoulos, {\em Impact of an Elastic Pseudo-Rigid Body
on a Rigid Foundation}, Int. J. Engrg. Sci., {\bf 38}, pp. 589--603, (2000).
\bibitem{Sol1999} 
J. M. Solberg and P. Papadopoulos, {\em A Simple Finite Element-Based Frame-
work for the Analysis of Elastic Pseudo-rigid Bodies}, Int. J. Num. Meth.
Engrg., {\bf 45}, pp. 1297--1314, (1999).
\bibitem{Sop1975} 
D. E. Soper, {\em Classical Field Theory}, John Wiley \& Sons, 1975.
\bibitem{Ste2000} 
I. E. Stepanova, A.V. Shchepetilov, {\em The Two-Body Problem on Spaces of
Constant Curvature. II. Spectral Properties of The Hamiltonian}, Teoret Mat
Fizika {\bf 124} (3) (2000), pp. 481--489.
\bibitem{Stoj1957} 
R. Stojanovitch, On the dynamics of a rigid body in Riemannian spaces,
ZAMM, 37, no.{\bf 7/8}, (1957).
\bibitem{Tul1962} 
B. Tulczyjew, W. M Tulczyjew, {\em On Multipole Formalism in General Relativity,Recent Developments in General Relativity}. Warsaw: Polish Scientific
Publishers, 1962., p. 465
\bibitem{Vill2005} 
R. Villanueva, M. Epstein, {\em Vibrations of Euler's Disk}, Phys. Rev. E71,
066609, 2005.
\bibitem{Woz1992} 
C. Wo\'{z}niak, {\em Mechanics of Continuous Media}, in Foundations of Mechanics, ed. H. ZorskiPWN, Warsaw, 1992.
\bibitem{Woz1988} 
C. Wo\'{z}niak, {\em Constraints in Mechanics of Continuous Bodies}, PAN-
Publisher, Wroc\l aw-Warszawa-Krak\'{o}w-Gda\'{n}sk-\L \'{o}d\'{z}, 1988
\bibitem{Wullf2002} 
C. Wullf, M. Roberts, {\em Hamiltonian Systems Near Relative Periodic Orbits}, SIAM Journal on Applied Dynamical Systems, {\bf 1}, 1--43 (2002).
\bibitem{Yav2008} 
A. Yavari, J. E. Marsden,{\em Covariant Balance Laws in Continua with Microstructure}, Accepted to Reports on Mathematical Physics, (2008).
\bibitem{Zie2005} 
O. C. Zienkiewicz, R. L. Taylor, J. Z. Zhu, {\em The Finite Element
Method},Elsevier LTD, Oxford, (2005).
\bibitem{Zor1967} 
M. Zorawski, {\em Th\'{e}orie Math\'{e}matique des Dislocations} Dunod, Paris (1967).
\end{thebibliography}
\end{document}